\documentclass[aps,prd,groupedaddress,showpacs,showkeys]{revtex4}
\usepackage{epsfig}
\pdfoutput=1
\newcommand{\be}[1]{\begin{equation} \label{(#1)}}
\newcommand{\eq}{\begin{eqnarray}}
\newcommand{\ee}{\end{equation}}
\newcommand{\en}{\end{eqnarray}}
\newcommand{\ba}[1]{\begin{eqnarray} \label{(#1)}}
\newcommand{\ea}{\end{eqnarray}}
\newcommand{\nn}{\nonumber}
\newcommand{\rf}[1]{(\ref{(#1)})}

\begin{document}

\title{Sterile neutrinos in lepton number and lepton flavor violating decays}
\author{Juan Carlos Helo,  Sergey Kovalenko, Ivan Schmidt  \vspace*{0.3\baselineskip}}
\affiliation{Universidad T\'ecnica Federico Santa Mar\'\i a, \\
Departamento de F\'\i sica, \\
and
Centro-Cient\'\i fico-Tecnol\'{o}gico de Valpara\'\i so, \\
Av. Espa\~na 1680, Valpara\'\i so,  Chile \vspace*{0.3\baselineskip}\\}

\date{\today}

\begin{abstract}
We study the contribution of massive dominantly sterile neutrinos, $N$,  to
the Lepton Number and Lepton Flavor Violating semileptonic decays of $\tau$ and  $B, D, K$-mesons. We focus on special
domains of sterile neutrino masses $m_{N}$ where it is close to its mass-shell. This leads to an enormous resonant 
enhancement of the decay rates of these processes. This allows us to
derive stringent limits on the sterile neutrino mass $m_{N}$ and its mixing $U_{\alpha N}$ with active flavors.
We apply a joint analysis of the existing experimental bounds on the decay rates of the studied processes. In contrast to other approaches in the literature our limits are free from ad hoc assumptions on the relative size of the sterile neutrino mixing parameters.  We analyze the impact of this sort of assumptions on the extraction of the limits on 
$m_{N}$ and $U_{\alpha N}$, and discuss the effect of finite detector size. Special attention was paid to the limits on meson decays with 
$e^{\pm}e^{\pm}$ in final state, derived from non-observation of $0\nu\beta\beta$-decay.  We point out that observation of these decays 
may, in particular,  shed light on the Majorana phases of the neutrino mixing matrix.
\end{abstract}



\maketitle  

\newpage

\section{Introduction}
\label{sec-1}

Neutrino oscillation experiments have shown conclusively 
that neutrinos are massive, although very light, particles mixing with each other. 
Moreover, neutrino oscillations is the first and so far  the only observed phenomenon 
of lepton flavor violation (LFV). Theoretically LFV is only possible if neutrinos are massive. 
Being electrically neutral neutrinos can be 
either Dirac or Majorana type particles. Majorana neutrino
masses violate total lepton number conservation and, therefore, can induce lepton number violating (LNV) processes.
Up to date such processes have not yet been observed experimentally.  Searching for LNV processes 
is a challenging quest, pointed to probe the nature of neutrinos and 
answer the question about wether they are
Majorana or Dirac particles.  However it  is well known that in 
Standard Model (SM) extensions with
very light and very heavy neutrinos, the rates of both LFV and LNV processes with charged leptons are  so small
that their experimental observation turns out to be unrealistic. Perhaps the only lucky exception is 
neutrinoless double beta decay ($0\nu\beta\beta$). The experiments searching for this LNV  process
are believed to reach even such small neutrino masses \cite{dbd-probe, KlapdorKleingrothaus:2000sn} as those which are relevant for neutrino oscillations,
and to probe the heavy Majorana neutrino sector up to the masses of several TeV \cite{Bamert:1995,Benes:2005hn}.

The situation may dramatically change, if there exist either moderately heavy fermions mixed with 
the active flavors $\nu_{e,\mu,\tau}$ or if there are some new LFV and LNV interactions beyond the SM. 
In this kind of extensions of the SM both LFV and LNV effects in the charged lepton sector could become significant.

Here we study the case of moderately heavy neutrinos in the scenario with $n$ species
of SM singlet right-handed neutrinos
$\nu^{\,\prime}_{R\alpha}=(\nu^{\,\prime}_{R1},...\nu^{\,\prime}_{Rn})$,
besides the three left-handed weak doublet neutrinos
$\nu^{\,\prime}_{L\beta} =
(\nu^{\,\prime}_{Le},\nu^{\,\prime}_{L\mu},\nu^{\,\prime}_{L\tau})$ \cite{Valle:1980,tau-decPaper}.
The general mass term for this set of fields can be written as 
\begin{eqnarray}\label{Mass-Term}
-\frac{1}{2} \overline{\nu^{\,\prime}} {\cal M}^{(\nu)} \nu^{\,\prime
c} + \mbox{h.c.} & = & - \frac{1}{2} (\overline{\nu^{\,\prime}}_{_L},
\overline{\nu_{_R}^{\,\prime c}}) \left(\begin{array}{cc}
{\cal M}_L & {\cal M}_D \\
{\cal M}^T_D  & {\cal M}_R \end{array}\right) \left(\begin{array}{c}
\nu_{_L}^{\,\prime c} \\
\nu^{\,\prime}_{_R}\end{array}\right) + \mbox{h.c.} \\
&=&-\frac{1}{2} (\sum_{i=1}^{3} m_{\nu_i} \overline{\nu^c_i}\nu_i
+\sum_{j=1}^{n} m_{\nu_j} \overline{\nu^c_j}\nu_j  )+ \mbox{h.c.}
\end{eqnarray} 
Here ${\cal M}_L, {\cal M}_R$ are $3\times 3$ and $n\times n$
symmetric Majorana mass matrices, and ${M}_D$ is a $3\times n$ Dirac
type matrix. Rotating the neutrino mass matrix to the diagonal form by a unitary
transformation
 \ba 
 && U^T {\cal M}^{(\nu)}U =
\textrm{Diag}\{m_{\nu 1}, \cdots ,m_{\nu_{3+n}}\}
\label{mass-eigen}
\ea
one ends up with $3+n$
Majorana neutrinos with masses $m_{v_1}, \cdots, m_{v_{3+n}}$.  The matrix $U_{\alpha k}$ is a neutrino mixing matrix.
In special cases among neutrino mass eigenstates there may appear  pairs with masses degenerate in
absolute values. Each of these pairs can be collected into a Dirac neutrino
field. This situation corresponds to conservation of certain lepton numbers
assigned to these Dirac fields. Generically in this setup neutrino mass eigenstates can be of any mass. 
For consistency with neutrino phenomenology (for a recent review, cf.  \cite{rev-nu-phen}) among them there must be the three very light neutrinos with different masses and  dominated by the active flavors $\nu_{\alpha}$ ($\alpha = e, \mu, \tau$).  The remaining states, conventionally called heavy sterile neutrinos, may also contain certain admixture of the active flavors and, therefore, participate in charged and neutral current interactions of the SM contributing to LNV and LFV processes. 
Explanation of the presence in the neutrino spectrum of  the three very light neutrinos requires additional physically motivated assumptions on the structure of the mass matrix in (\ref{Mass-Term}). The celebrated ``see-saw'' mechanism \cite{see-saw}, presently called type-I see-saw, is implemented in this framework assuming that  $\mathcal{M}_R\gg \mathcal{M}_D$. Then,  there naturally appear light neutrinos with masses of the order of $\sim\mathcal{M}_D^2/\mathcal{M}_R$ dominated by $\nu_{\alpha}$. Also, there must be present heavy Majorana neutrinos with masses at the scale of  $\sim \mathcal{M}_R$. Their mixing with active neutrino flavors is suppressed by a factor $\sim \mathcal{M}_D/\mathcal{M}_R$ which should be very small. In particular scenarios this generic limitation of the see-saw mechanism can be relaxed \cite{relax}, \cite{Kersten:2007vk}, \cite{Petcov:2010}. Then the heavy sterile neutrinos could be, in principle, observable at LHC, if their masses are within the kinematical reach of the corresponding experiments. Very heavy or moderately heavy Majorana entry 
$\mathcal{M}_R$ of the 
neutrino mass matrix naturally appears in various extensions of the SM. The well known examples are given by the $SO(10)$-based supersymmetric \cite{SUSY-GUT} and  ordinary \cite{GUT} grand unification models as well as models with spontaneous breaking of lepton number \cite{Valle:1982}.
The supersymmetric versions of see-saw  are also widely discussed in the literature 
(see, for instance,  \cite{SUSY-see-saw} and references therein). 

In the present paper we study the above mentioned generic case of the neutrino mass matrix in (\ref{Mass-Term}) without implying a specific scenario of neutrino mass generation.  
We assume there is one moderately heavy sterile neutrino $N$ in the MeV-GeV mass domain.
The presence or absence of these neutrino states
is a question for experimental searches.
If exist, they may contribute to some LNV and LFV processes as intermediate nearly on-mass-shell states. This would lead to enormous  resonant enhancement of their contributions to these processes.  As a result, it may become possible to either observe the LNV, LFV processes or  set stringent limits on sterile neutrino mass $m_{N}$  and mixing $U_{\alpha N}$ with active neutrino flavors $\nu_{\alpha}$ ($\alpha = e, \mu, \tau$) from non-observation of the corresponding processes.

On the other hand the sterile neutrinos in this mass range are motivated by various phenomenological models \cite{Mohapatra}, in particular, by the recently proposed electroweak scale see-saw models \cite{EWSS1},  \cite{EWSS2}.  
They may also play an important astrophysical and cosmological role (for a recent review see, for instance,  \cite{Kusenko:2009up}).  The sterile neutrinos in this mass range may have an impact on Big Bang nucleosynthesis, large scale structure formation \cite{nuclsyn}, supernovae explosions \cite{supernovae}. Moreover, the keV-GeV sterile neutrinos are good dark matter candidates \cite{DM-Bar-1,DM-Bar-2,DM-Bar-3}  and offer a
plausible explanation of  baryogenesis \cite{Barg}.
Dark Matter sterile neutrinos, having small admixture of active flavors, may suffer radiative decays and contribute to the diffuse extragalactic radiation and x-rays from galactic clusters \cite{galclast}. 
This is, of course, an incomplete list of cosmological and astrophysical  implications of sterile neutrinos.  More details on this subject can be found  in  Refs. \cite{Dolgov1}, \cite{Smirnov1}.

The phenomenology of sterile neutrinos in the processes, which can be searched for in laboratory experiments have been studied in the literature in different contexts and from complementary points of view (for earlier studies see \cite{Shrock}).
Their resonant contributions to $\tau$ and meson decays have been studied in Refs.  
\cite{K-decPaper, tau-decPaper,Ivanov:2004ch, Cvetic:2010rw,Atre:2009rg}. 
Another potential process to look for  sterile Majorana neutrinos is like-sign dilepton production in hadron collisions  
\cite{Almeida:2000pz,Panella:2001wq,Han:2006ip,Kovalenko:2009td}.  Possible implications of sterile neutrinos have been also studied in LFV muonium decay and high-energy muon-electron scattering \cite{Cvetic:2006yg}.
Constraints on the sterile neutrino parameters have been derived from the accelerator and Super-Kamiokande measurements  \cite{Kusenko:2004qc}.
For a recent review of sterile neutrino phenomenology we refer readers to Ref. \cite{Atre:2009rg}. 
An interesting explanation of anomalous excess of events observed in the LSND \cite{LSND} and MiniBooNE \cite{MiniBooNE} neutrino  experiments has been recently proposed  \cite{Gninenko} in terms of  sterile neutrinos with masses from 40 MeV to 80 MeV.  An explanation comes out of their possible production in neutral current interactions of $\nu_{\mu}$ and subsequent radiative decay  to light neutrinos.

The present paper contributes to some still uncovered aspects of the phenomenology of sterile neutrinos.  
We focus on the derivation of limits on the sterile neutrino mixing matrix elements $|U_{e N}|, |U_{\mu N}|, |U_{\tau N}|$
from the experimental data on LFV and LNV decays of $K, D, B$-mesons and $\tau$. In our analysis we use no 
ad hoc assumptions on the relative size of  these matrix elements, typically used in the similar studies existing  in the literature. 
We examine an impact of this sort of assumptions on the resulting limits for $|U_{\alpha N}|$. 

The paper is organized as follows. 
In Sec. \ref{sec-2} we present decay rate formulas for $\tau$ and meson LFV and LNV decays and their reduced forms, in the resonant domains of the sterile neutrino mass.
In Sec. \ref{sec-3} we discuss theoretical uncertainties in the calculation of the total decay width of sterile neutrinos and present our ``inclusive'' approach based on Bloom-Gilman duality. This approach allows one to avoid uncertainties related to the heavy meson decay constants, which are  typical for the conventional channel-by-channel approach.  
In Sec. \ref{sec-4} we discuss the extraction of the limits on  the active-sterile neutrino mixing matrix elements 
$|U_{e N}|, |U_{\mu N}|, |U_{\tau N}|$
and present the corresponding exclusion plots. 
In Sec. \ref{sec-5} we discuss the $0\nu\beta\beta$-decay constraints on the sterile neutrino parameters and their possible implications for probing Majorana phases of  the neutrino states. Here we also derive predictions for the rates of some LNV and LFV decays of $K, D, D_{s},B, B_{c}$-mesons as yet unconstrained experimentally. 
Sec. VI summarizes our main results.

\section{Decay rates}
\label{sec-2}

Neutrino interactions are represented by the SM Charged (CC) and Neutral Current (NC) Lagrangian terms.  In the mass eigenstate basis they read
\begin{eqnarray}\label{CC-NC}
{\cal L} = \frac{g_2}{\sqrt{2}}\sum_{i}\  U_{l i}\ \bar l \gamma^{\mu} P_L \nu_i\  W^-_{\mu}
+  \frac{g_2}{2 \cos\theta_W}\  \sum_{\alpha, i, j}U_{\alpha j} U_{\alpha i}^*\  \bar\nu_i \gamma^{\mu} P_L \nu_j\  
Z_{\mu},
\end{eqnarray}
where $l = e,\mu,\tau$ and $i=1,..., n+3 $. We consider the case with a single sterile neutrino $N$ with a mass $m_{N}$ and, therefore,  in (\ref{CC-NC}) we choose $n=1$ and identify $N=\nu_{4}$.

\begin{figure}[htbp]
\centering
\includegraphics[width=0.8\textwidth,bb=100 580 540 730]{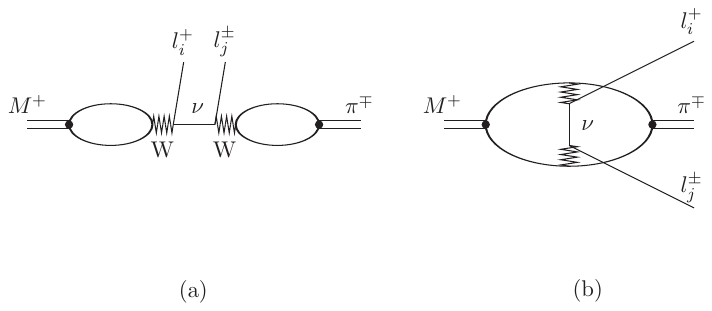}
\caption{The lowest order diagrams contributing to the semileptonic  meson decays. } 
\label{fig-1}
\end{figure}

In what follows we analyze the above mentioned resonant contribution of heavy sterile neutrino to the semileptonic LNV and LFV   decays of $\tau$ and the pseudoscalar mesons $M = K, D, B$:
\ba{BD}
\tau^-\rightarrow l^{\mp} \pi^{\pm} \pi^-, \ \  \ \ M^+\rightarrow l_i^+  l_j^{\pm}\pi^{\mp} .
\ea
Lowest order diagrams for the case of meson decays are shown in Fig. \ref{fig-1}. There is only one tree-level diagram with an intermediate Majorana or Dirac neutrino, shown in Fig. \ref{fig-1}(a),
contributing to LFV decays. For LNV decays mediated by Majorana neutrinos,
in addition to the tree-level diagram Fig. \ref{fig-1}(a), there appears a two-loop diagram shown in Fig. \ref{fig-1}(b).
The tree-level diagram is known \cite{Ivanov:2004ch} to dominate in the processes  \rf{BD}. As will be seen, 
in the studied domain of the sterile neutrino mass, the two-loop diagram in Fig. \ref{fig-1}(b) is absolutely negligible. 
An important point is that the calculation of the tree-level diagram, Fig.  \ref{fig-1}(a),
does not require knowledge of the hadronic structure needed for the diagram in Fig. \ref{fig-1}(b).

\newpage
The decay rates of the studied processes are given by the expressions:
\ba{rate-Lv}
 &LNV:&  \Gamma(M^+\rightarrow l_i^+ l_j^+ \pi^- )= \nn \\
 &&
=\kappa_{ij} \int\limits_{s_{i}^-}^{s_{j}^+} d s \sum_{k} \left | \frac{U_{i k}U_{j k} m_{\nu k}}{s - m_{\nu k}^2}\right |^2 G^{i j}_{M}(\frac{s}{m_{_M}^2})
+ \kappa_{ij} \int\limits_{s_{j}^-}^{s_{i}^+} d s \sum_{k} \left | \frac{U_{i k}U_{j k} m_{\nu k}}{s - m_{\nu k}^2}\right |^2 G^{ji}_{M }(\frac{s}{m_{_M}^2}) +  \nn \\
&&+ 4\kappa_{ij}  {\rm Re}[\int\limits_{s_i^-}^{s_j^+} d s_1
\left( \frac{U_{i k}U_{j k} m_{\nu k}}{s_1 - m_{\nu k}^2}\right)
\int\limits_{s_{ij}^-}^{s_{ij}^+} d s_2 \left(\frac{U_{i k}U_{j k} m_{\nu k}}{s_2 - m_{\nu k}^2}\right)^* H^{i j}_{ M}(\frac{s_1}{m_{_M}^2},\frac{s_2}{m_{_M}^2})],\\
\label{rate-Lfv}
 &LFV: & \Gamma(M^+\rightarrow  l_i^+ l_j^- \pi^+)=  
\int\limits_{s_{i}^-}^{s_{j}^+} d s \sum_{k} \left | \frac{U_{i k}U^{*}_{j k}}{s - m_{\nu k}^2}\right |^2 
s\  G^{i j}_{M}(\frac{s}{m_{_M}^2}).
 \ea
Here the unitary mixing matrix $U_{ij}$ relates $\nu'_i = U_{ij}\nu_j$,
the weak $\nu'$ and mass $\nu$ neutrino eigenstates.
In Eq. \rf{rate-Lv} the factor $\kappa_{ij}=1-\delta_{ij}/2$  takes into account a combinatorial factor 1/2 for
identical final leptons. 
In the above equations we introduced the following functions:
\ba{G-funct}
 G^{i j}_{M}(z)& = & c^{M} \frac{\phi^{ij}(z)}{z^{2}}
\left[h_{i-}^2(z)-x_\pi^2 h_{i+}(z) \right]\left[h_{j+}(z)-
h_{j-}^2(z)\right],
\\ \nn
H^{i j}_{M}(z_1,z_2)&=& \frac{c^{M}}{m_{_M}^{2}} \left\{\frac{}{} h_{i-}(z_1)h_{j-}(z_2)+ x_\pi^2
\left[h_{i-}(z_1)h_{j-}(z_2)+x_i^2+x_j^2\right]- \right.\\ &&
 \left.\frac12
\left[h_{i-}(z_1)+h_{j-}(z_2)\right]
\left[h_{i-}(z_1)h_{j+}(z_2)+h_{i+}(z_1)h_{j-}(z_2)\right]\right\}.\nn
\ea
with
\ba{phi-1}
&&\phi^{ij}(z)= \lambda^{1/2}(z,x_{i}^2,x_{\pi}^2)\lambda^{1/2}(z,x_{j}^2,1), \ \ \ \ h_{i\pm}(z)=z\pm x^2_{i}\, ,\\
\label{cM}
&&c^{M} \ \ \ \ =  \frac{G_{F}^{4}}{128 \pi^{3}}  f^2_{\pi}  f^2_{_M} m^5_{_M} |V_{ud}|^2|V_{M}|^2,\ \ \ \lambda(x,y,z) = x^2+y^2+z^2-2xy -2yz- 2xz\, ,
\ea
and the dimensionless variables $x_{i} = m_{l_i}/m_{_M}$, $x_{\pi} = m_{\pi}/m_{_M}$,
where $m_{l_i}$ and $m_{_M}$ are the masses of the charged leptons $l_i=e, \mu, \tau$ and the initial meson $M$ respectively.
Numerical values of meson masses $m_{M}$,  decay constants  $f_{_M}$ and the CKM factors $V_{M}$
are specified in Table I.

\begin{table}[h!]  
\begin{center}
{{\bf Table I:} Masses $m_{P,V}$ and decay constants $f_{P,V}$ of pseudoscalar P and vector V mesons.}\\[5mm]
\begin{tabular}{|c|c|c|c||c|c|c|c|}\hline
${\bf P}$& $m_{P}$ (MeV) \cite{PDG} & $f_{P}$ (MeV)& $V_{P}$ & ${\bf V}$ & $m_{V}$ (MeV) \cite{PDG} & $f_{V}$ (MeV) & $V_{V}$\\
\hline
$\pi^{\pm}$  & 139.6 & 130.7 \cite{PDG}& $V_{ud}$ & $\rho^{\pm}$ &775.8 & 220 \cite{Ebert:2006hj} & $V_{ud}$\\
\hline
$K^{\pm}$ & 493.7 & 159.8  \cite{PDG} & $V_{us}$ & $K^{*\pm}$ & 891.66 & 217 \cite{Ebert:2006hj} &$V_{us}$\\
\hline
$D^{\pm}$ &1869.4 &222.6 \cite{CLEO:2005} & $V_{cd}$ & $D^{*\pm}$ & 2010 & 310 \cite{Ebert:2006hj} &$V_{cd}$\\
\hline
$D^{\pm}_{s}$ & 1968.3 & 266 \cite{PDG} & $V_{cs}$ & $D^{*\pm}_{s}$ & 2112.1 & 315 \cite{Ebert:2006hj} &$V_{cs}$\\
\hline
$B^{\pm}$ & 5279 & 190 \cite{MILC:2002} &    $V_{ub}$ & $ \rho^{0}$ & 776& 220 \cite{Ebert:2006hj} &- \\
\hline
$B^{\pm}_{c}$ & 6277 & 399 \cite{Ivanov:2006ni} & $V_{cb}$ & $\omega$& 782.59 & 195 \cite{Ebert:2006hj} & -\\
\hline
$\pi^{0}$ & 135 & 130\cite{PDG}& - & $K^{*0}, \bar{K}^{*0}$ & 896.1 & 217 \cite{Ebert:2006hj} & -\\
\hline
$K^{0}, \bar{K}^{0}$ & 497.6 & 159  \cite{PDG} & - &  $\phi$ & 1019.456 & 229 \cite{Ebert:2006hj} & -\\
\hline
$\eta$ & 547.8 & 164.7  \cite{Feldmann: 2000} & - &  $D^{*0}, \bar{D}^{*0}$& 2006.7 & 310 \cite{Ebert:2006hj}& -\\
\hline
$\eta'$ & 957.8 & 152.9 \cite{Feldmann: 2000} & - &  $J/\psi$  & 3096.916 & 459 \cite{Wang:2006}& -  \\
\hline
$\eta_{c}$ & 2979.6 & 335.0 \cite{CLEO:2001}& -  &   &  &  &   \\
\hline
$B^{0}_{s}$ & 5367.5 & 216 \cite{Cvetic:2004qg}& - &   &  &  &   \\
\hline
\end{tabular}
\end{center}
\end{table}

The integration limits in Eqs. \rf{rate-Lv}, (\ref{rate-Lfv}) are
\ba{lim-int}
&&s_{l}^- = m_{_M}^2(x_{\pi} + x_{l})^2, \ \ \ \ s_{l}^+ = m_{_M}^2(1
- x_{l})^2,\\ \nn &&s_{l_1l_2}^{\pm} =\frac {m_{_M}^2}{2y}
\left[(x_{l_1}^2-x_{\pi}^2)(x_{l_2}^2-1)+y(1+x_{l_1}^2+x_{l_2}^2+x_{\pi}^2)-
y^2\pm \phi^{l_1l_2}(y) \right] \ea with $y=s_1/m_{_M}^2$.

If one assumes that neutrinos are separated into light $\nu_k$
and heavy $N_k$ states, with masses $m_{\nu i} << \sqrt{s_l^-}$
and $\sqrt{s_l^+} << m_{N_{k}}$,  then the branching ratios \cite{K-decPaper} turn out to be extremely small:
 \ba{limit11}
 R_{l_i l_j} = \frac{\Gamma(M^{+}\rightarrow l_{i}^{+} l_{j}^{+} \pi^{-})}{\Gamma({\rm
M}^+\rightarrow all)} \leq \ \sim 10^{-30} \ \ \ 3\ light\
neutrino\ scenario. 
\ea
\ba{limit12}
 R_{l_i l_j} \leq\ \sim 10^{-19}\ \ \ 3\ light
+ 1\ heavy\ neutrino\ scenario
\ea
These values are far beyond experimental reach.

On the other hand, if we assume that there exists at least one massive neutrino $N$
with mass $m_N$ in the range  
\ba{domain}
min\left[ \sqrt{s_{i}^-}, \sqrt{s_{j}^-}\right]\leq m_{N} \leq
max\left[\sqrt{s_{i}^+}, \sqrt{s_{j}^+}\right]
\ea
the situation changes drastically.
Then the s-channel M-decay
diagram in Fig \ref{fig-1}(a)  blows up because the integrands under the single
integrals in Eqs. \rf{rate-Lv}, (\ref{rate-Lfv}) have a
non-integrable singularity at $s =m_N^2$ corresponding to an on-mass-shell intermediate neutrino. Therefore, in this
resonant domain the total  decay width of heavy neutrino
$\Gamma_{N}$ has to be taken into account. This can be done by
the substitution $m_{N}\rightarrow m_{N} - (i/2)\Gamma_{N}$. 
As will be seen in sec. \ref{sec-4} the heavy neutrino width  $\Gamma_{N}$ is very small in the resonant domain, $\Gamma_{N}\approx 10^{-9}$ MeV, 
\cite{K-decPaper,tau-decPaper}. Therefore,  the neutrino propagator in the single integrals of
Eqs. \rf{rate-Lv}, (\ref{rate-Lfv}) has a very sharp maximum at
$s=m_{N}^2$. The double integrals, being finite in the limit
$\Gamma_{N} =0$, can be neglected in the considered case.
Thus, with good precision we obtain from Eqs.\rf{rate-Lv}, (\ref{rate-Lfv}) the following decay rate formulas for a nearly 
on-mass-shell intermediate neutrino:
\begin{eqnarray}\label{LNV-res}
 LNV: \ \Gamma^{res}(M^{+}\rightarrow \pi^{-} l^{+}_{i}l^{+}_{j})& \approx& \kappa_{ij}
\pi (G_M^{ij}(x_{N}^{2})+G_M^{ji}(x_{N}^{2}))\frac{m_N |U_{i N}|^2|U_{j N}|^2} {\Gamma_{N}} \\
\label{LFV-res} 
 LFV: \ \Gamma^{res}(M^{+}\rightarrow \pi^{+} l^{-}_{i}l^{+}_{j})& \approx&
\pi G_M^{ij}(x_{N}^{2})\frac{m_N |U_{i N}|^2|U_{j N}|^2} {\Gamma_{N}},
\end{eqnarray}
with 
$x_{N} = m_{N}/m_{M}$.
On the other hand these formulas follow directly from the fact that in the resonant domain a heavy neutrino, produced in meson decay $M\rightarrow l_{i} N$, propagates as a real unstable particle
which then decays to $N\rightarrow l_{j} \pi$.  In the narrow-width approximation these two processes are independent and one can write
\begin{eqnarray}\label{nwa}
\Gamma^{res}(M^{+}\rightarrow l^{+}_{i}l^{+}_{j} \pi^{-} )& \approx& \kappa_{ij} \left(\Gamma(M^{+}\rightarrow l_{i}^{+} N)\frac{\Gamma(N\rightarrow l_{j}^{+} \pi^{-})}{\Gamma_{N}} + 
i \leftrightarrow j\right)
\\
\label{nwa1}
\Gamma^{res}(M^{+}\rightarrow  l^{+}_{i}l^{-}_{j} \pi^{+})& \approx& 
\Gamma(M^{+}\rightarrow l_{i}^{+} N)\frac{\Gamma(N\rightarrow l_{j}^{-} \pi^{+})}{\Gamma_{N}},
\end{eqnarray}
where meson decay rates are \cite{Shrock}
\begin{eqnarray}\label{Mes}
\Gamma(M^{+}\rightarrow l_{i}^{+} N) = |U_{i N}|^{2} \frac{G_{F}^{2}}{8\pi} f_{M}^{2} |V_{M}|^{2} m_{M}^{2}
\sqrt{\lambda(x_{i}^{2},x_{N}^{2},1)}(x_{i}^{2} + x_{N}^{2} - (x_{i}^{2} - x_{N}^{2})). 
\end{eqnarray}
This expression exhibits  proper chiral suppression behavior and vanishes in the limit $m_{l}, m_{N} = 0$ as a result of the weak charged current V-A structure  and angular momentum conservation. Since $m_{N}$ is large, this process is not suppressed even for electrons in the  final state of the decay $M\rightarrow e N$, in contrast to the well known examples of chirally suppressed leptonic decays of pseudoscalar mesons  such as $\pi^{-}\rightarrow e^{-}\bar{\nu}_{e}$. 
An expression for $\Gamma(N\rightarrow l \pi)$ is given below, in Eq. (\ref{lP}).  Substituting these formulas into Eqs. (\ref{nwa}), (\ref{nwa1}) one can easily reproduce Eqs. (\ref{LNV-res}), (\ref{LFV-res}). 

Similar arguments apply to tau decays $\tau^- \rightarrow l^+ \pi^- \pi^+$ \cite{tau-decPaper}, which 
in the  neutrino mass domain \mbox{$m_{\pi}+m_{l} \leq m_N \leq m_{_\tau} - m_{\pi}$}, lead also to resonantly 
enhanced decay rates:
\begin{eqnarray}\label{tau-dec} 
\Gamma^{res}(\tau^- \rightarrow l^\pm \pi^\mp \pi^-) &\approx&  
\Gamma(\tau^{-} \rightarrow \pi^{-} N)\frac{\Gamma(N\rightarrow l^{\pm}\pi^{\mp})}{\Gamma_{N}}= 
\pi G^{l}(z_{N}^{2})\frac{m_N |U_{\tau N}|^2|U_{l N}|^2} {\Gamma_{N}}
\end{eqnarray}           
 where
 \ba{G-funct2} 
 G^{l}(z)& = &  c^{\tau}\frac{\phi^{l}(z)}{z^{2}}
\left[(z-z_l^2)^2-z_\pi^2(z+z_l^2) \right]\left[ (z-1)^2-z_\pi^2(z+1)\right], \nn \\
\phi^{l}(z) & = & \lambda^{1/2}(z,z_{l}^2,z_{\pi}^2)\lambda^{1/2}(z,z_{\pi}^2,1), \ \ \ 
c^{\tau} = \frac{G_{F}^{4}}{128 \pi^{3}} f^4_{\pi}   m^5_{\tau} |V_{ud}|^4
\ea
with  $z_{N} =  m_{N}/m_{\tau}$, $z_{l} = m_{l}/m_{\tau}$, $z_{\pi} = m_{\pi}/m_{\tau}$. 

We will use the above decay rate formulas Eqs. (\ref{LNV-res}), (\ref{LFV-res}) and (\ref{tau-dec}), \rf{G-funct2} in our analysis of the sterile neutrino contribution to $\tau$ and meson decays.

\section{Heavy Sterile Neutrino Decay Rate}
\label{sec-3}
Heavy sterile neutrinos $N$,  being mixed with active neutrino flavors, can decay into various final states depending on their mass 
$m_{N}$. 
These decays are represented by purely leptonic 
$N\rightarrow l_{1}l_{2} \nu, \ 3\nu$ and semileptonic $N\rightarrow l {\cal H}$ modes, where ${\cal H}=M, B,..$ are hadronic states represented by mesons and baryons. These decays proceed via 
CC and NC interactions of the SM given by the Lagrangian (\ref{CC-NC}).
The total  decay rate, $\Gamma_{N}$,  of heavy neutrinos is conventionally calculated in the literature in the {\it channel-by-channel} approach
\cite{K-decPaper, Atre:2009rg}, in which one sums up the partial decay rates of all the leptonic and two-body semileptonic  decay channels open for a given  value of $m_{N}$.  
In the present paper we apply another approach. This is the {\it inclusive} approach we proposed in 
Ref. \cite{tau-decPaper}, in which we approximate the semileptonic decays of the heavy neutrino $N$ by its decays into quark-antiquark pairs $N \rightarrow l(\nu) q_1 \bar q_2$, as suggested  by Bloom-Gilman duality \cite{BloomGilman}.  This implies that in an average over sufficiently wide range of the invariant mass of the final hadronic state ${\cal H}$, the sum of all the open decay channels $N\rightarrow l(\nu) {\cal H}$ is approximately equal to the rate of $N \rightarrow l(\nu) q_1 \bar q_2$.
In comparison with the former approach the latter does not require information on the parameters of the final mesons, such as masses and decay constants, some of which are poorly known for the meson states starting from  $\eta'(985)$ 
and heavier. The inclusive approach is supposed, according to the duality arguments, to take into account all the semileptonic channels and, therefore,  in this case $\Gamma_{N}$ should be larger than in the channel-by-channel approach, in which some hadronic states are neglected. We apply a simplified version of the inclusive approach, neglecting  
perturbative and nonperturbative QCD corrections to the tree-level quark production diagram. This leading-order approximation is expected to be reasonable for $m_{N}>> \Lambda \approx 200$ MeV. At lower masses a more viable approach 
would be to relate  by dispersion relations the semileptonic $N$ decay rate
to the imaginary parts of the W and Z self-energies $\Pi(s)$, in analogy to the approach
applied in the literature for the $\tau\rightarrow \nu +  hadrons$ inclusive decay
\cite{SelfEnergy}.
However for our rough estimations we do not need this more sophisticated treatment and
will use the above mentioned leading-order approximation.

Here we summarize the partial decay rates for the inclusive approach \cite{tau-decPaper}, including leptonic and semileptonic decay modes of the heavy sterile neutrino N.   In the latter case the final hadronic states for low neutrino masses $m_{N}$ is represented by the lightest mesons while for larger $m_{N}$ by $q\bar{q}$-pairs.  The list of decay rates is as follows:
\begin{eqnarray}\label{lln-CC}
\Gamma(N\rightarrow l_1^{-}l_2^{+}\nu_{l_{2}} )&=& |U_{l_1 N}|^2
\frac{G_F^2}{192\pi^3} m_N^5 I_{1}(y_{l_1},y_{\nu_{l_{2}}}, y_{l_2})(1-\delta_{l_{1}l_{2}}) 
\equiv |U_{l_1 N}|^2 \Gamma^{(l_1l_2\nu)}, \\
\label{lln}
\Gamma(N\rightarrow \nu_{l_{1}}l_2^{-}l_2^{+} )&=& |U_{l_1 N}|^2
\frac{G_F^2}{96\pi^3} m_N^5
\left[\left(g^{l}_{L} g^{l}_{R}+ \delta_{l_{1}l_{2}}g^{l}_{R}\right) I_{2}(y_{\nu_{l_{1}}}, y_{l_{2}}, y_{l_{2}}) + \right. \\ \nn
&&\left.   + \left((g^{l}_{L})^{2} +(g^{l}_{R})^{2 }+ \delta_{l_{1}l_{2}} (1 +2 g^{l}_{L})\right) I_{1}(y_{\nu_{l_{1}}}, y_{l_{2}}, y_{l_{2}}) \right] \equiv
\\ \nn
&\equiv& |U_{l_1 N}|^2\Gamma^{(l_2l_2\nu)},\\
\label{3n}
\sum_{l_{2}=e,\mu,\tau}\Gamma(N\rightarrow \nu_{l_{1}} \nu_{l_{2}} \bar{\nu}_{l_{2}})&=& |U_{l_1 N}|^2 
\frac{G_F^2}{96\pi^3} m_N^5  \equiv |U_{l_1 N}|^2
\Gamma^{(3\nu)},\\
\label{lP}
\Gamma(N\rightarrow l^{-}_{1} P^{+}) &=& |U_{l_{1}N}|^2
\frac{G_F^2}{16 \pi}m_N^3 f_{P}^2 |V_{P}|^{2} F_P(y_{l_{1}},y_{P})\equiv |U_{l_{1}N}|^2
\Gamma^{(lP)},\\
\label{nuP}
\Gamma(N\rightarrow \nu_{l_{1}} P^0) &=& |U_{l_{1}N}|^2 \frac{G_F^2}{64 \pi}m_N^3  f_{P}^2 
(1 - y^2_{P})^2 \equiv |U_{l_{1}N}|^2 \Gamma^{(\nu P)},\\ 
\label{lV}
\Gamma(N\rightarrow l^{-}_{1} V^{+}) &=& |U_{l_{1}N}|^2 
\frac{G_F^2}{16\pi}m_N^3 f_{V}^2 |V_{V}|^{2} F_V(y_{l_{1}},y_{V})\equiv |U_{l_{1}N}|^2
\Gamma^{(lV)},\\ 
\label{nuV}
\Gamma(N\rightarrow \nu_{l_{1}} V^0) &=&  |U_{l_{1}N}|^2 \frac{G_F^2}{2 \pi}m_N^3 \ f_{V}^2\  \kappa_{V}^{2} (1 - y^2_{V})^2
(1 + 2 y^2_{V}) \equiv |U_{l_{1}N}|^2 \Gamma^{(\nu V)},\\
\label{lud}
\Gamma(N\to l_{1}^{-} u\bar{d})&=&|U_{l_{1}N}|^2\ |V_{u d}|^{2} \frac{G_F^2}{64\pi^3}m_N^5 
I_{1}(y_{l_{1}}, y_{u}, y_{d})\equiv |U_{l_{1}N}|^2\Gamma^{(lud)},\\
 \label{nuqq}
\Gamma(N\to \nu_{l_{1}}\, q\bar{q}) &=& |U_{l_{1}N}|^2\frac{G_F^2}{32\pi^3}m_N^5F_{Z}(m_{N}) \left[ g^{q}_{L} g^{q}_{R} I_{2}(y_{\nu_{l_{1}}}, y_{q}, y_{q}) + \right. \\ \nn
&&\left. + \left((g^{q}_{L})^{2} +(g^{q}_{R})^{2 })\right) I_{1}(y_{\nu_{l_{1}}}, y_{q}, y_{q}) \right] \equiv |U_{l_1 N}|^2 
\Gamma^{(\nu qq)}.
\end{eqnarray} 
Here we denoted $y_{i} = m_{i}/m_{N}$ with $m_{i} = m_{l}, m_{P}, m_{V}, m_{q}$. For the quark masses we use the values 
\mbox{$m_{u}\approx m_{d} = 3.5$ MeV},  \mbox{$m_{s} = 105$ MeV}, \mbox{$m_{c} = 1.27$ GeV}, \mbox{$m_{b} = 4.2$ GeV.}  In Eqs. (\ref{lud}), (\ref{nuqq})  we denoted  $u=u,c,t$; $d=d,s,b$ and $q=u,d,c,s,b,t$. The SM neutral current couplings of leptons and quarks are
\begin{eqnarray}\label{NC-coupl}
g^{l}_{L} &=&-1/2 + \sin^2\theta_W, \  \  g^{u}_{L}= 1/2 - (2/3) \sin^2\theta_W,  \ \ 
 g^{d}_{L}= -1/2 + (1/3) \sin^2\theta_W, \\
 \nn
g^{l}_{R} &=& \sin^2\theta_W,\ \ \ \ \ \ \ \ \ \ \  \ \,  g^{u}_{R}= -(2/3)\sin^2\theta_W,\  \ \ \ \  \ \ \,  g^{d}_{R}= (1/3)\sin^2\theta_W,
\end{eqnarray}
The corresponding NC couplings of mesons are given by
\begin{eqnarray}\label{NC-mesons}
\kappa_{V} &=& \sin^2\theta_W/3\ \ \ \ \ \ \ \ \ \ \  \ \, \mbox{for}\ \  \rho^{0}, \omega, \\
\nn
\kappa_{V} &=& -1/4 + \sin^2\theta_W/3 \ \ \mbox{for} \ \ K^{*0}, \bar{K}^{*0}, \phi, \\
\nn  
\kappa_{V} &=& 1/4 - 2 \sin^2\theta_W/3\ \  \, \mbox{for}\ \  D^{*0}, \bar{D}^{*0}, J/\psi
\end{eqnarray}

The kinematical functions are
\ba{kin-fun-1}
 &&I_{1}(x,y,z)= 12 \int\limits_{(x+y)^{2}}^{(1-z)^{2}} \frac{ds}{s}
(s-x^2-y^{2})(1+z^2-s) \lambda^{1/2}(s, x^{2}, y^2) \lambda^{1/2}(1, s, z^2),\nn
\\ 
&& I_{2}(x,y,z)= 24 y z \int\limits_{(y+z)^{2}}^{(1-x)^{2}} \frac{d s}{s} (1+x^{2}-s)
\lambda^{1/2}(s,y^{2},z^{2})\lambda^{1/2}(1,s,x^{2}), \\ \nn
&&F_P(x,y)= \lambda^{1/2}(1,x^2,y^2) [(1+x^2)(1+x^2-y^2) - 4 x^2],\nn \\ \nn
&&F_V(x,y)= \lambda^{1/2}(1,x^2,y^2) [(1-x^2)^2+(1+x^2)y^2 - 2 y^4],\\ \nn
\ea

\begin{figure}[htbp]
\centering
\includegraphics[width=0.8\textwidth]{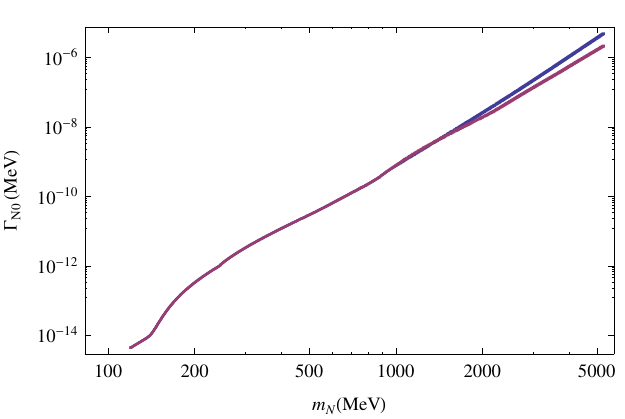}
\caption{Majorana sterile neutrino decay rate calculated in the inclusive (upper curve) and channel-by-channel approaches for $|U_{eN}|=|U_{\mu N}|=|U_{\tau N}| = 1$.}
\label{fig-2}
\end{figure}

The total decay rate $\Gamma_{N}$  of the heavy neutrino $N$ is equal to the sum of the partial decay rates in Eqs. (\ref{lln-CC})-(\ref{nuqq}),
which we write in the form: 
\ba{total-4}
\Gamma_{N}& =&  \sum_{l_{1}, l_{2}, {\cal H}}\left[ \eta_{N}  \Gamma(N\rightarrow l_{1}^{-}{\cal H}^{+}) +
\eta_{N}  \Gamma(N\rightarrow l_{1}^{-} l_{2}^{+}\nu_{l_{2}}) +\right. \\ \nn
&+&\left.\Gamma(N\rightarrow \nu_{l_{1}}{\cal H}^{0}) + \Gamma(N\rightarrow l_{2}^{-} l_{2}^{+}\nu_{l_{1}}) + 
\Gamma(N\rightarrow \nu_{l_{1}}\nu_{l_{2}} \bar{\nu}_{l_{2}}) \right],
\ea
where we denoted the hadronic states ${\cal H}^{+} = P^{+}, V^{+}, \bar{d} u,  \bar{s} u,  \bar{d} c,  \bar{s} c$ and  ${\cal H}^{0} = P^{0}, V^{0}, \bar{q} q$. We introduced the factor $\eta_{N} = 2$ for Majorana and $\eta_{N}=1$ for Dirac neutrino $N$. Its value 
$\eta_{N}=2$ is related with the fact that for Majorana neutrinos both charge conjugate final states are allowed: 
$N\rightarrow l_{1}^{-}l_{2}^{+} \nu_{l_{2}}, l_{1}^{+}l_{2}^{-} \bar{\nu}_{l_{2}}$ and $N\rightarrow l^{\mp}{\cal H}^{\pm}$. 
For convenience we write Eq. \rf{total-4} in the form:
\ba{GT-comp-incl}
\Gamma_{N} &=& a_{e}(m_{N})\cdot  |U_{eN}|^{2}  + a_{\mu}(m_{N})\cdot   |U_{\mu N}|^{2} +  
a_{\tau}(m_{N})\cdot   |U_{\tau N}|^{2}
\ea
where 
\begin{eqnarray}\label{coeff-a}
a_{l}(m_{N}) = 
\Gamma^{(l {\cal H})} + \Gamma^{(3\nu)} +\sum_{l_{2}} \left(\Gamma^{(l_{2}l_{2}\nu)} + \eta_{N} \Gamma^{(l_{1} l_{2} \nu)}\right),
\end{eqnarray}
and $l, l_{2} = e, \mu, \tau$. In the  inclusive approach the hadronic contribution is calculated as
\begin{eqnarray}\label{H-incl}
\Gamma^{(l{\cal H})} = \theta(\mu_{0}-m_{N})\sum_{P,V}\left(\Gamma^{(\nu P)} + \Gamma^{(\nu V)}  + \eta_{N} \Gamma^{(lP)} 
+ \eta_{N} \Gamma^{(lV)}\right) +
\theta(m_{N} - \mu_{0})\sum_{u,d,q} \left(\eta_{N} \Gamma^{(lud)} + \Gamma^{(\nu qq)}\right)
\end{eqnarray}
In Eqs. (\ref{coeff-a}) and (\ref{H-incl}) we used notations for $\Gamma^{(ijk)}$ and $\Gamma^{(ij)}$ introduced in Eqs. (\ref{lln-CC})-(\ref{nuqq}). The parameter $\mu_{0}$ denotes the mass threshold from which we start taking into account hadronic contributions via 
$q \bar{q}$ production. In our numerical study we use the mass $\mu_{0} = m_{\eta'} = 957.8$ MeV. 
At this rather low threshold, $\mu_{0}$, the QCD corrections become significant, but  it is reasonable to expect that the tree-level contribution still dominates the semileptonic decay rates since $\mu_{0}>\Lambda_{QCD}$. We believe that the accuracy of the above presented inclusive approach is sufficient for estimations of limits on the parameters of sterile neutrino in this range of masses. In this respect it is worth noticing that  in the channel-by-channel approach uncertainties related to the heavy meson decay constants $f_{M}$ and other hadronic parameters are large and theoretically not well controllable. 
As seen from the references in \mbox{Table I}, 
many of decay constants $f_{P,V}$ are only known in phenomenological models. 
In order to compare our inclusive approach with the channel-by-channel one, we show in Fig. \ref{fig-2}  the heavy neutrino total decay rate 
$\Gamma_{N}(m_{N})$, calculated in both approaches
assuming $|U_{eN}|=|U_{\mu N}|=|U_{\tau N}| = 1$.
As seen, the inclusive curve, which starts from $m_{N}=\mu_{0}=957.8$ MeV,  gradually deviates upward from the channel-by-channel one. This tendency becomes more pronounced for large $m_{N}$. This gradual deviation is explained by the fact that the inclusive approach takes into account in average 
all the possible decay channels of the heavy neutrino. As a result the decay rate in this approach is larger. Some of these channels cannot be taken into account in the channel-by-channel approach since some heavy resonances, especially with b- and c-quarks, are poorly known or not yet known at all. In the literature based on the channel-by-channel approach, these channels are disregarded. 

\section{Limits on Sterile Neutrino Mass and Mixing}
\label{sec-4}

In the literature there are various limits on the mass $m_{N}$ and mixing $U_{\alpha N}$ (with $\alpha = e,\mu, \tau$) of  a sterile neutrino $N$,
extracted from direct and indirect experimental searches \cite{PDG} for this particle, in a wide region of its mass. A recent summary of these limits, extracted from the corresponding 
experimental data, is given in Ref. \cite{Atre:2009rg}.  
In Figs. \ref{fig-3}-\ref{fig-4} we show the exclusion plots from Ref. \cite{Atre:2009rg} together with our exclusion curves, 
derived in a model independent way on the basis of a joint analysis of semileptonic LNV and LFV decays of $K, B, D, D_{s}$-mesons and $\tau$. Our approach is discussed in what follows.

\subsection{Joint Analysis}
\label{sec-4-1}
As seen from Eqs.  (\ref{LNV-res}),  (\ref{LFV-res}), (\ref{tau-dec}) and \rf{GT-comp-incl}, the decay rates of these processes  depend on all the three  mixing matrix elements $U_{\alpha  N}$ with $\alpha = e,\mu,\tau$. In the literature it is a common practice to adopt some ad hoc assumptions on their relative size in order to extract limits on them from the existing experimental bounds on the  corresponding  decay rates. These  assumptions reduce the reliability of the obtained limits. Our method is based on a joint analysis of the above mentioned processes, without any additional ad hoc assumptions of this sort. 
We apply a numerical Monte Carlo sampling in a parametric space 
\begin{eqnarray}\label{sample}
&&|U_{eN}|, |U_{\mu N}|, |U_{\tau N}|, m_{N} \ \ \ \mbox{with}\ \ \  
m_{N} \in Res(M,\tau)\\
\label{unitarity}
&&  \mbox{and} \ \ \ |U_{eN}|^{2}+|U_{\mu  N}|^{2}+|U_{\tau N}|^{2}\leq \Delta
\end{eqnarray}
taking into account the existing experimental bounds on the branching ratios of LNV and LFV decays: \mbox{$M\rightarrow ll \pi$}, 
\mbox{$\tau\rightarrow l \pi\pi$}. Here $Res(M,\tau)$ denotes resonant regions of sterile neutrino mass corresponding 
to the meson \mbox{$M=K,B,D,D_s$} and $\tau$ decays included in the analysis. 
The experimental limits on their branching ratios were taken from Ref. \cite{PDG}, and  are shown in Table II (without brackets).
The constraint (\ref{unitarity}) originates from the unitarity of the $4\times 4$ neutrino mixing matrix 
$ |U_{eN}|^{2}+|U_{\mu  N}|^{2}+|U_{\tau N}|^{2}+ |U_{s N}|^{2}=1$, where the sterile neutrino admixture,  $|U_{s N}|^{2}$, is model dependent and unlimited experimentally.
We used in our analysis $\Delta = 1$, which leads to the most relaxed limits on $|U_{e N}|, |U_{\mu N}|, |U_{\tau N}|$.
We checked that these limits remain unaffected for smaller values of this cutoff parameter such that  $\Delta\geq 0.1$.
Thus, despite the parameter $|U_{\tau N}|$ is unconstrained by the experimental data its value $|U_{\tau N}|^{2}\leq 0.1$
is compatible with our limits discussed below. Note that the value of this mixing matrix element should not be very close
to 1. This follows from the fact that the unitarity of the mixing matrix  
$|U_{\tau 1}|^{2}+|U_{\tau 2}|^{2}+|U_{\tau 3}|^{2}+|U_{\tau N}|^{2} = 1$ in the limit $|U_{\tau N}|^{2}=1$
leads   to $|U_{\tau 1}|^{2}+|U_{\tau 2}|^{2}+|U_{\tau 3}|^{2} = 0$ inconsistent with the light neutrino phenomenology.

\begin{figure}[htbp]
\centering
\includegraphics[width=0.8\textwidth]{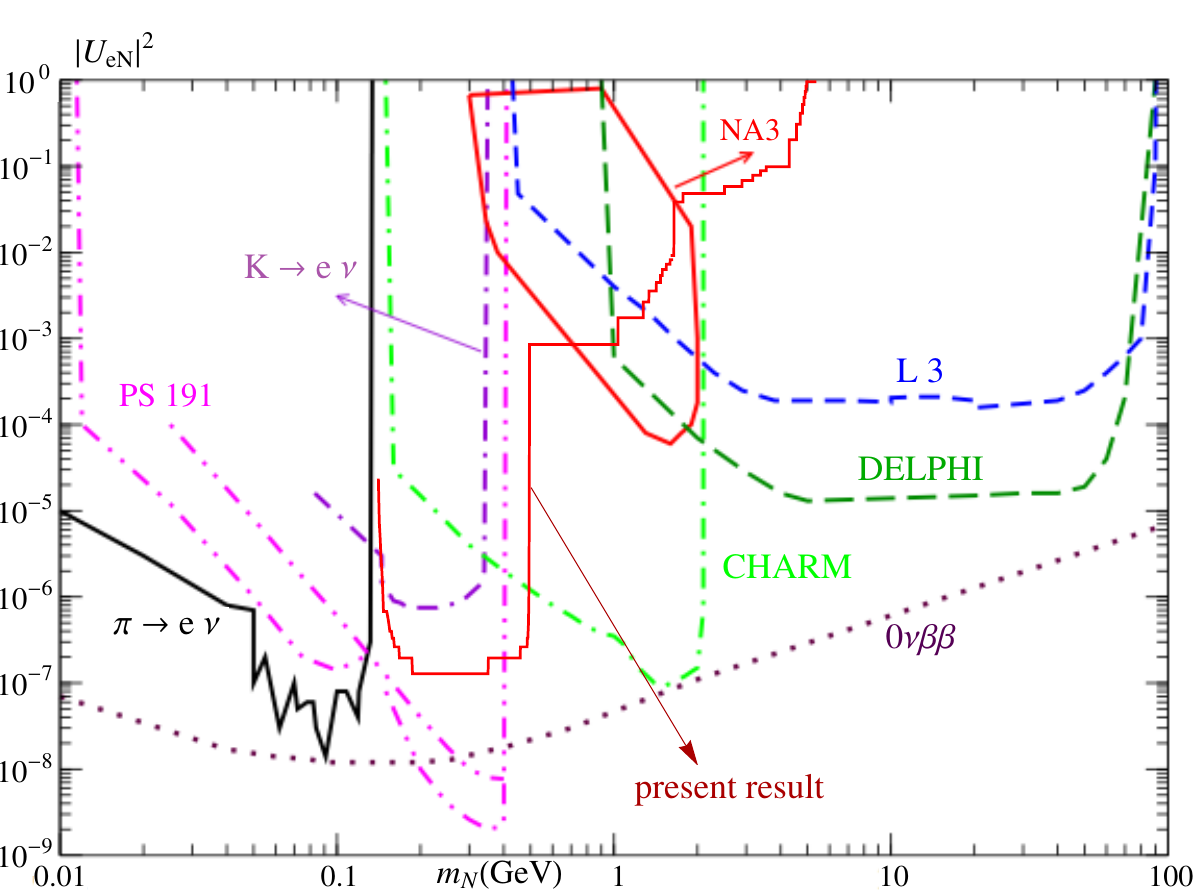}
\caption{Exclusion plots for the mixing matrix element $|U_{eN}|^2$ from various experimental searches \cite{PDG}. The corresponding curves are taken from Ref. \cite{Atre:2009rg}. 
The solid line marked ``present result'' is our exclusion curve derived from a joint analysis of the existing experimental upper bounds \cite{PDG} on the rates of  $\tau$ and $K, D, B$-meson semileptonic LNV and LFV decays.}
\label{fig-3}
\end{figure}

\begin{figure}[htbp]
\centering
\includegraphics[width=0.8\textwidth]{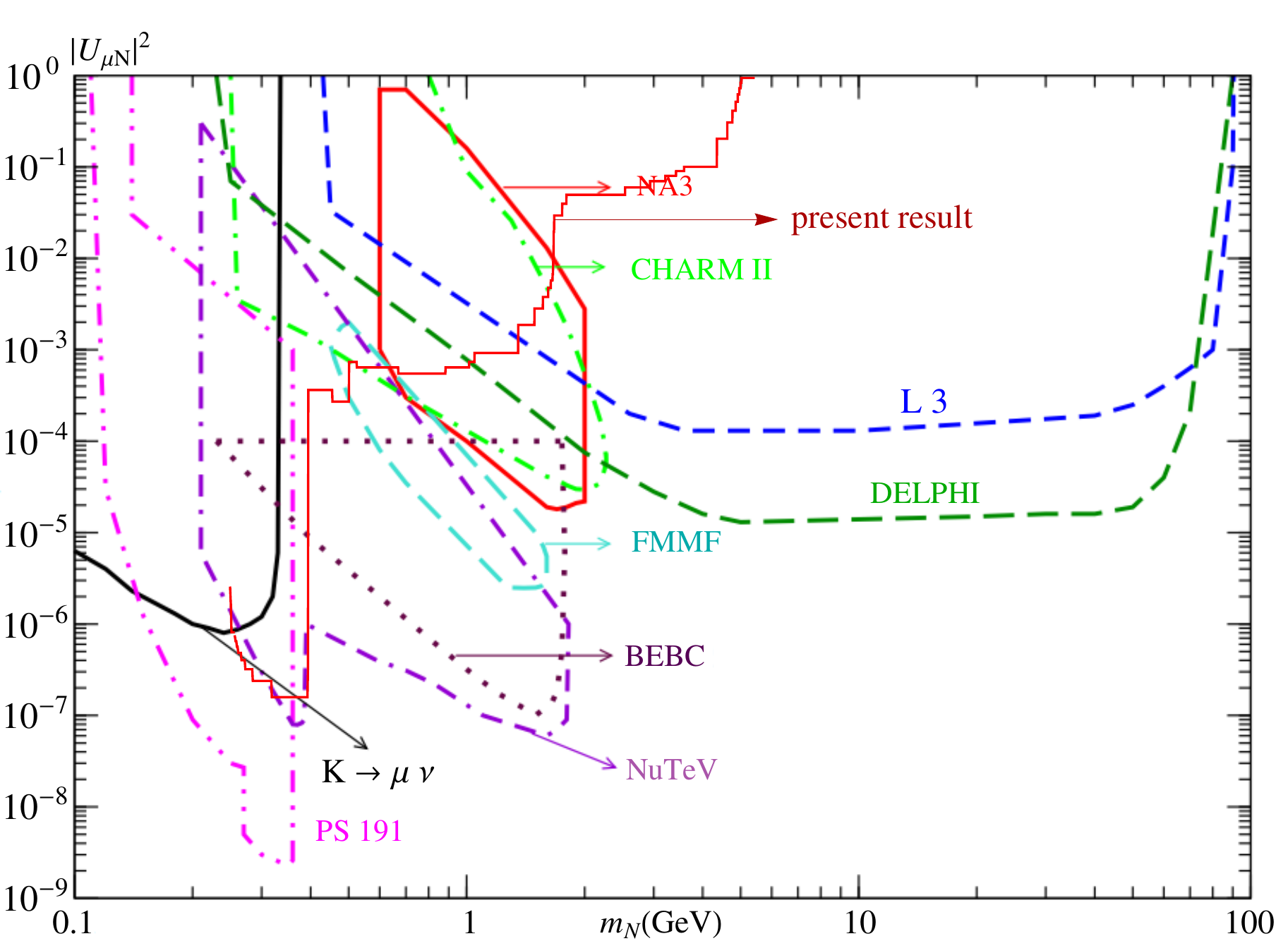}
\caption{The same as in Fig. \ref{fig-3} but for $|U_{\mu N}|^2$. The existing exclusion curves   
are taken from Ref. \cite{Atre:2009rg}. }
\label{fig-4}
\end{figure}

\begin{figure}[htbp]
\centering
\includegraphics[width=0.6\textwidth]{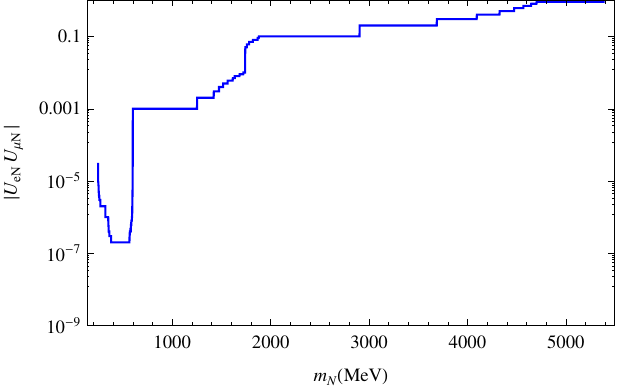} 
\includegraphics[width=0.6\textwidth]{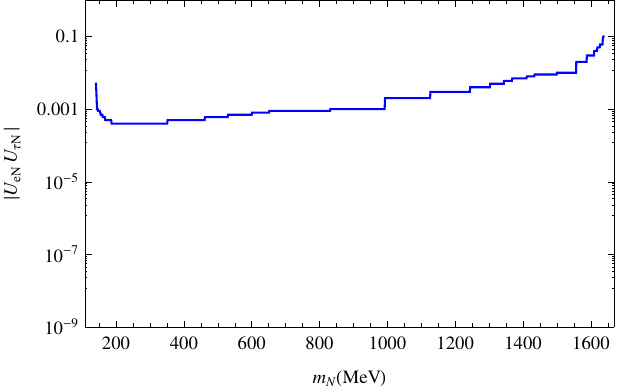}
\includegraphics[width=0.6\textwidth]{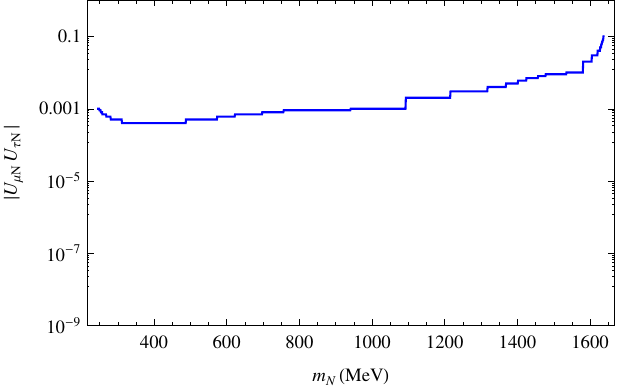}
\caption{Exclusion curves for products of mixing matrix elements
derived from a joint analysis of the existing experimental bounds \cite{PDG} on the rates of  $\tau$ and $K, D, B$-meson semileptonic LNV and LFV decays.}
\label{fig-5}

\end{figure}

The obtained exclusion curves for $|U_{eN}|$ and $|U_{\mu N}|$ are shown in Figs. \ref{fig-3}, \ref{fig-4}, together with other existing limits. 
In Figs. \ref{fig-5} we  present our exclusion curves for $|U_{eN} U_{\mu N}|$, $|U_{eN} U_{\tau N}|$ and $|U_{\mu N} U_{\tau N}|$. Here we do not show
limits from other processes \cite{Atre:2009rg} since in the region of our interest, $m_{N}\leq 5$GeV,  they are significantly less stringent than ours.

The following comment is in order. The $3\times 3$ sub-block of the neutrino mixing matrix $U_{\alpha k}$ in 
Eqs. (\ref{sample})-(\ref{unitarity}) corresponds to the mixing of the light dominantly active neutrinos. As is known this sector of neutrino mixing is well constrained from the neutrino oscillation experiments (for a recent review see, for instance, Ref. \cite{global-fit}). However, most of the constraint of this type have been obtained in the three light neutrino mixing scenario with the unitary $3\times 3$ mixing  matrix. In the present context this implies no mixing with the sterile neutrino: 
$U_{e N}= U_{\mu N}=U_{\tau N} = 0$.  Recently there have also been analyzed implications of non-unitary light neutrino mixing matrix including scenarios with one and two light sterile neutrino states (see, for instance, 
Refs. \cite{Antusch:2006vwa,Goswami:2008mi,Rodejohann:2009ve,Kopp:2011qd} and references therein). In this case neutrino oscillation data impose rather week limits on $|U_{\alpha N}|$ of the order of $< 10^{-1}$,  which are significantly weaker than our limits and the other limits  presented in Figs.  \ref{fig-3}-\ref{fig-5}. For this reason we do not show them here.

Let us point out that our exclusion curves in Figs. \ref{fig-3}-\ref{fig-5} represent the best possible limits which can be  extracted from the existing experimental bounds on LNV and LFV semileptonic decays of $\tau$ and mesons shown in Table II
without ad hoc assumptions on the relative size of the mixing  matrix elements.  However certain assumption of this type may be physically reasonable and will be discussed in subsection \ref{sec-4-4}.
 
\subsection{Finite Detector Size Effect}
\label{sec-4-2}
The following important comment is now in order.   In the studied case of
the resonantly enhanced  LFV and LNV decays the sterile neutrino $N$  is close to its mass shell. Thus, as we already mentioned at the end of sec. \ref{sec-2},  the sterile neutrino, produced in $\tau$ and mesons  
decays 
$M\rightarrow l N$ and $\tau\rightarrow \pi N$, 
propagates as a real particle and decays at a certain distance from the production point. 
If this distance is larger than the size of the detector, the neutrino escapes from it before decaying, and the signature of 
$\tau\rightarrow l\pi\pi$ or $M_{1}\rightarrow ll M_{2}$ cannot be recognized. As a result no limits on $m_{N}$ and $|U_{iN}|$ can be extracted from the experimental upper bounds on the branching ratios of these processes. The limits we derived above correspond to the idealized case of infinitely large detector. 
In the realistic case the finite detector size effect must be taken into account.  Its impact on the results of the data analysis essentially depends on concrete experimental setup.
In particular, the finite size of detector implies taking into account the probability,  ${\cal P}_{N}$, of heavy neutrino decay within a detector. It should be incorporated in the analysis of concrete experimental data and can only be made by the corresponding experimental group. In general, the probability of a particle, in our case a sterile neutrino, decay within a detector of the length $L_{D}$ is given by
\begin{eqnarray}\label{PN-1}
{\cal P}_{N}=1-exp\left(-L_{D}\Gamma_{N} \frac{m_{N}}{p_{N}}\right),
\end{eqnarray}
where $p_{N}$ and $\Gamma_{N}$ are the  sterile neutrino 3-momentum and the total decay rate. The latter was calculated in Sec. \ref{sec-3}. 
The sterile neutrino is produced in decays $M\rightarrow l N$ and $\tau\rightarrow \pi N$. Therefore, its momentum, $p_{N}$, crucially depends on the energy spectrum of the initial mesons and $\tau$. 
In the rest frame of the initial mesons and $\tau$ 
the sterile neutrino momentum $\tilde{p}_{N}$ has the following value fixed by the sterile neutrino mass:
\begin{eqnarray}\label{restfr}
\tilde{p}_{N} = \frac{1}{2 m_{A}} \left((m_{A}^{2} -m_{B}^{2} - m_{N}^{2})^{2} -4m_{N}^{2}m_{B}^{2} \right)^{1/2},
\end{eqnarray}
where $m_{A}=m_{M}$, $m_{B}=m_{l}$ for the decays $M\rightarrow l N$ and $m_{A}=m_{\tau}$, $m_{B}=m_{\pi}$ for $\tau\rightarrow \pi N$.  

In Ref. \cite{Atre:2009rg}  the probability factor ${\cal P}_{N}$ was considered under an inconsistent assumption that the sterile neutrino is relativistic while its gamma factor is $\gamma=1$.  
We do not apply this assumption. In order to illustrate the impact of the effect of finite detector size we assume that the initial mesons and $\tau$ in decays \rf{BD} are at rest or very slow. In this case the 3-momentum $p_{N}$ of the intermediate on-mass-shell sterile neutrino   in Eq.  (\ref{PN-1}) equals to $\tilde{p}_{N}$ given by (\ref{restfr}). 
Then for the extraction of limits on $|U_{iN}|$, taking into account the finite detector size, we use the following condition
\begin{eqnarray}\label{LD}
\Gamma^{th} {\cal P}_{N} \leq \Gamma^{exp},
\end{eqnarray}
where 
$\Gamma^{th}$  are  given in Eqs. (\ref{LNV-res})-(\ref{tau-dec}) and 
$\Gamma^{exp}$ are experimental upper bounds on the corresponding decay rates. The probability factor 
${\cal P}_{N}$, given in Eq. (\ref{PN-1}),  depends on the ratio $\tilde{p}_{N}/m_{N}$. As follows from (\ref{restfr}) this ratio vanishes at the upper border of the resonant regions 
\rf{domain} and reaches   
its maximal value at the lower border.
Thus the effect of the probability factor ${\cal P}_{N}$ have significant impact on (\ref{LD}) only in the region of low values of $m_{N}$ and 
$|U_{iN}|$. In this region we write Eq. (\ref{PN-1}) approximately 
${\cal P}_{N}\approx L_{D}\Gamma_{N} m_{N}/\tilde{p}_{N}$.  Following Ref. \cite{Atre:2009rg} we assume for simplicity 
$|U_{eN}|=|U_{\mu N}|=|U_{\tau N}|$ and obtain from  
Eq. (\ref{LD}) a rough estimate for the limits corrected by the finite detector size
\begin{eqnarray}\label{Ueff}
|U_{iN}|^{2} = \sqrt{\frac{|U_{iN}|^{2}_{\infty}}{L_{D} \Gamma_{N0}}\frac{\tilde{p}_{N}}{m_{N}}},
\end{eqnarray}
where $|U_{iN}|_{\infty}$ is a limit obtained assuming infinite detector size and $\Gamma_{N0}$ is the sterile neutrino decay rate, calculated according to Eq. \rf{GT-comp-incl} with $|U_{eN}|=|U_{\mu N}|=|U_{\tau N}| = 1$. 
We point out that  we use Eq. (\ref{Ueff}) only for illustrative purpose considering very slow initial mesons and $\tau$ in decays \rf{BD}. In this case we find that  the probability factor ${\cal P}_{N}$ affects only the parts of our exclusion curves for 
$|U_{e N}|^{2}$ and $|U_{e N}U_{\mu N}|$ within $140$ MeV$<m_{N}<400$ MeV in \mbox{Figs. \ref{fig-3}, \ref{fig-5}} and for 
$|U_{\mu N}|^{2}$  within $245$ MeV$<m_{N}<380$ MeV in Fig. \ref{fig-4}. 
As seen from Figs. \ref{fig-3}-\ref{fig-5}, typical idealized limits in these regions are 
$|U_{eN}|^{2}_{\infty}\sim |U_{\mu N}|^{2}_{\infty}<10^{-7}$. Then, as follows from Eq. (\ref{Ueff}), the finite detector size corrections with $L_{D}=10$ m make these limits significantly weaker: roughly by a factor $10^{-3}$, $10^{-2}$, $10^{-1}$ for 
$m_{N}=140$ MeV, 240 MeV, 380 MeV, respectively.   
Detailed effect of the finite detector size as well as other experimental conditions depend on experimental setup and should be taken into account in analysis of concrete experimental data.  As we mentioned above the results will crucially depend on the energy spectrum of the initial mesons and $\tau$. 
This sort of analysis is beyond the scope  of the present paper 
and we limit ourselves to the above presented general discussion of the influence of the finite detector size on the limits
derived from the decays \rf{BD}.

\subsection{Analytical Methods of Extraction of Limits}
\label{sec-4-3}
It is sometimes useful to have analytic expressions for the limits on sterile neutrino mixing.  Such expressions for the case of  $|U_{e N}|^2$ and $|U_{\mu N}|^2$, without ad hoc assumptions on 
the relative size of different matrix elements $|U_{l N}|^2$, can be derived directly from Eqs.  (\ref{LNV-res}) -  (\ref{tau-dec}), \rf{GT-comp-incl} and the unitarity relations (\ref{unitarity}). Taking  $\Delta = 1$ in  (\ref{unitarity}) one finds conservately
\ba{Ec2}
&&|U_{eN}|^4-|U_{eN}|^2 F_{ee}(M) (a_e-a_{\tau}) -F_{ee}(M)a_\mu < 0\\
\nonumber
&&|U_{\mu N}|^4-|U_{\mu N}|^2 F_{\mu \mu}(M) (a_{\mu}- a_{\tau}) -F_{\mu \mu}(M)a_{e} < 0\ \ \
\ea
Solving these inequalities one gets limits
\ba{Cota1}
&&|U_{eN}|^2<\frac{1}{2}\left[F_{ee}(M)(a_e - a_{\tau}) + \sqrt{F^{2}_{ee}(M)(a_e-a_{\tau})^2 + 4 F_{ee}a_\mu}\right]\\
\label{Cota1-12}
&&|U_{\mu N}|^2<\frac{1}{2}\left[F_{\mu\mu}(M)(a_{\mu} - a_{\tau}) + \sqrt{F^{2}_{\mu\mu}(M)(a_{\mu}-a_{\tau})^2 + 4 F_{\mu\mu}a_{e}}\right]\ea
valid in the resonant  regions of decays $M\rightarrow ee \pi$ and $M\rightarrow \mu \mu \pi$, respectively.
We introduced notations
\begin{eqnarray}\label{Notat-22}
F_{\mu e}(M) = \frac{\Gamma^{exp}(M\rightarrow\mu e \pi) }{\pi m_{N} (G^{\mu e}(x^{2}_{N}) + G^{e\mu}(x^{2}_{N})}, \ \ \ 
F_{ll}(M) = \frac{\Gamma^{exp}(M\rightarrow ll \pi) }{\pi m_{N}  G^{ll}(x^{2}_{N})}, \ \
F_{l}(\tau) = \frac{\Gamma^{exp}(\tau\rightarrow l\pi\pi) }{\pi m_{N} G^{l}(z_{N}^{2})}\ \ \  
\end{eqnarray}
in agreement with Eqs. (\ref{LNV-res})-\rf{G-funct2}. Here the $\Gamma^{exp}$ are experimental upper bounds on  the rate of the indicated decays.  Eqs. \rf{Cota1}, (\ref{Cota1-12}) allow setting limits on $|U_{e N}|^{2}$
and $|U_{\mu N}|^{2}$, considering different LNV and LFV decays independently, and using the existing experimental bounds on their rates. 

Stronger limits  can be extracted from a joint analysis of certain sets of LNV and LFV decays, if these decays 
have nontrivial intersection of their resonant regions in 
$m_{N}$.  Then for values of $m_{N}$ in the intersection one can extract limits for $U_{l N}$ 
in an analytic form in the following way.
Let us consider, for instance, the set of LNV and/or LFV decays: $M\rightarrow \pi e \mu$, $M\rightarrow \pi ee$ and $\tau \rightarrow \pi \pi e$. Then from Eqs.  (\ref{LNV-res}) -  (\ref{tau-dec}) and \rf{total-4} we obtain 
\ba{Cota3}
&&\frac{ |U_{e N}|^4} {a_e |U_{e N}|^2+a_\mu |U_{\mu N}|^2 +a_\tau |U_{\tau N}|^2} < F_{ee}(M),\  \ \
 \frac{ |U_{e N}|^2|U_{\mu N}|^2} {a_e |U_{e N}|^2+a_\mu |U_{\mu N}|^2 +a_\tau |U_{\tau N}|^2} < F_{e\mu}(M),\\
 \nonumber
&& \frac{ |U_{e N}|^2|U_{\tau N}|^2} {a_e |U_{e N}|^2+a_\mu |U_{\mu N}|^2 +a_\tau |U_{\tau N}|^2}  < F_{e}(\tau)
 \ea
combining these Eqs. we find
\ba{Cota4}
|U_{e N}|^2 < a_e F_{ee}(M)+a_\mu F_{e\mu}(M)+a_\tau F_{e}(\tau)
\ea 
Analogously, for the set  $M\rightarrow \pi \mu \mu$, $M\rightarrow \pi e\mu$ and $\tau \rightarrow \pi \pi \mu$ we have 
\ba{Cota3-2}
&&\frac{ |U_{\mu N}|^4} {a_e |U_{e N}|^2+a_\mu |U_{\mu N}|^2 +a_\tau |U_{\tau N}|^2} < F_{\mu\mu}(M),\  \ \
 \frac{ |U_{e N}|^2|U_{\mu N}|^2} {a_e |U_{e N}|^2+a_\mu |U_{\mu N}|^2 +a_\tau |U_{\tau N}|^2} < F_{e\mu}(M),\\
 \nonumber
&& \frac{ |U_{\mu N}|^2|U_{\tau N}|^2} {a_e |U_{e N}|^2+a_\mu |U_{\mu N}|^2 +a_\tau |U_{\tau N}|^2}  < F_{\mu}(\tau)
 \ea
and find the limit
\ba{Cota5}
|U_{\mu N}|^2 < a_e F_{e\mu}(M)+a_\mu F_{\mu\mu}(M)+a_\tau F_{\mu}(\tau)
\ea 
Using in these formulas \rf{Cota4}, \rf{Cota5},  the experimental bounds on the rates of the corresponding decays \cite{PDG}, one can
analyze all the sets of decays of the above mentioned type for $\tau$ and $B,D, D_{s}, K$-mesons, and derive exclusion curves in $|U_{eN}|^{2}-m_{N}$ and $|U_{\mu N}|^{2}-m_{N}$ planes. 
The resulting exclusion curve for $|U_{eN}|^2$ is limited to the mass range $m_N =141-1637$ MeV, while for $|U_{\mu N}|^2$ to the range $m_N= 246-1637$ MeV. We do not present these curves since they are not very different in these mass regions from those derived in the Monte Carlo approach and shown in Figs.  \ref{fig-3}-\ref{fig-5}. 
Thus the above presented analytical formulas  \rf{Cota4}, \rf{Cota5}, considered in the corresponding regions of sterile neutrino mass $m_{N}$, 
can be treated as a reasonable alternative to a more involving Monte Carlo analysis.

\subsection{Limits with Additional Assumptions on the Mixing Matrix Elements}
\label{sec-4-4}

\begin{figure}[htbp]
\centering
\includegraphics[width=0.7\textwidth]{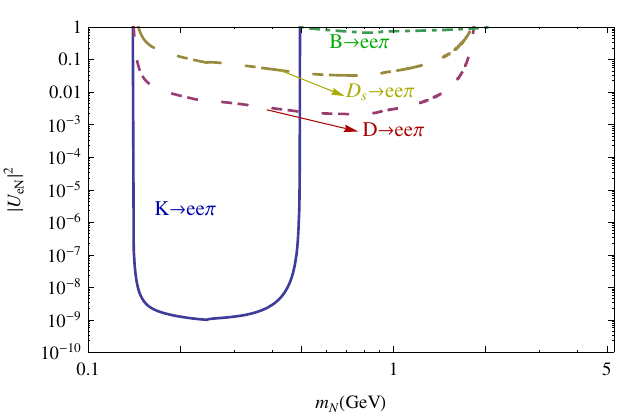}
\includegraphics[width=0.7\textwidth]{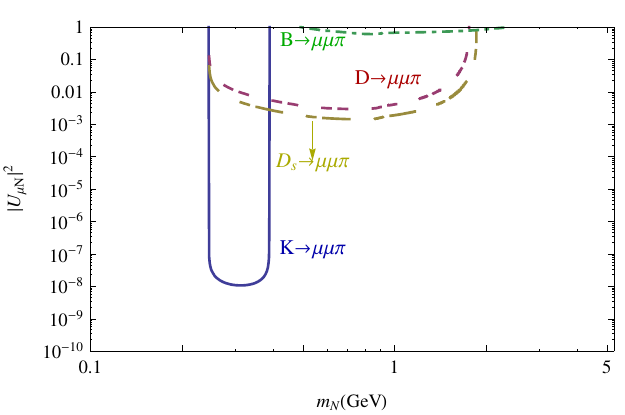}
\caption{Exclusion curves derived  with the assumption $|U_{eN}|\sim |U_{\mu N}|\sim |U_{\tau N}|$ from the experimental limits on LNV decay of mesons.}
\label{fig-6}
\end{figure}

\begin{figure}[htbp]
\centering
\includegraphics[width=0.6\textwidth]{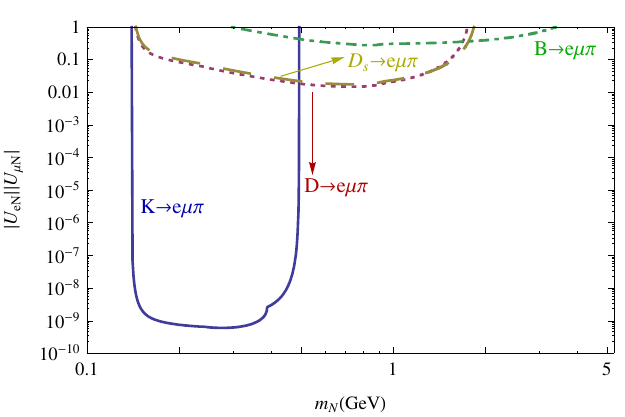}
\includegraphics[width=0.6\textwidth]{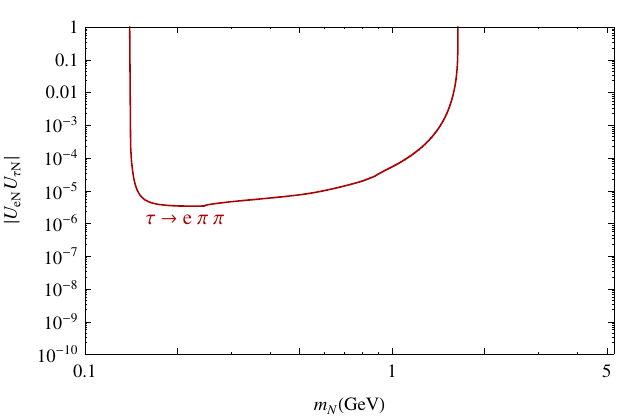}\\
\includegraphics[width=0.6\textwidth]{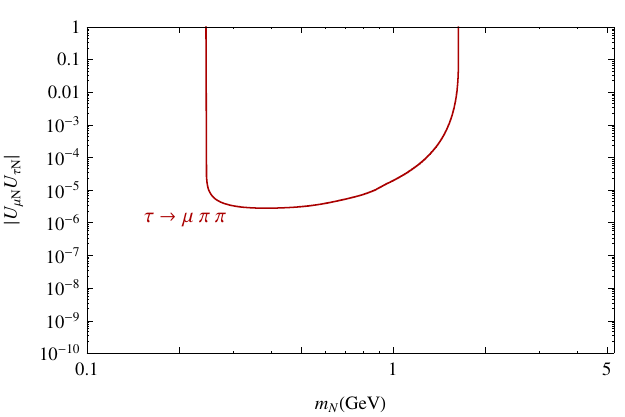}
\caption{Exclusion curves derived with the assumption $|U_{eN}|\sim |U_{\mu N}|\sim |U_{\tau N}|$ from the experimental limits on LNV decay of $\tau$ and mesons.}
\label{fig-7}
\end{figure}

In sec. \ref{sec-4-1} we extracted limits on $|U_{i N}|$ without any additional assumption on the relative size of these mixing matrix elements. 
However some assumptions may be physically reasonable and worth considering.  It can be noticed that in our 
joint analysis in sec.  \ref{sec-4-1} the size of $|U_{\tau N}|$ is not controlled by the experimental limits, because of absence of the corresponding experimental data on the LNV and LFV processes involving two $\tau$-leptons.   
In this situation some additional assumptions on the size of this and other mixing matrix elements, $|U_{i N}|$, look desirable, although they leave an imprint in the extracted limits making them less objective.  Let us consider the assumption 
$|U_{eN}|\sim |U_{\mu N}|\sim |U_{\tau N}|$, frequently used in the literature.  Then from Eqs. \rf{Cota3},  \rf{Cota3-2} one finds upper limits in an analytic form
\begin{eqnarray}\label{lim-a1}
&&|U_{e N}|^{2}\ \ \ \  < F_{ee}(M)(a_{e}+a_{\mu}+a_{\tau}); \ \ \  \  |U_{\mu N}|^{2} \ \ \ \ < F_{\mu \mu}(M)(a_{e}+a_{\mu}+a_{\tau}); \\
&& |U_{e N}U_{\mu N}| < F_{e\mu}(M)(a_{e}+a_{\mu}+a_{\tau}); \ \ \  
|U_{e N}U_{\tau N}| < F_{e}(\tau)(a_{e}+a_{\mu}+a_{\tau}); \\
&&|U_{\mu N}U_{\tau N}| < F_{\mu}(\tau)(a_{e}+a_{\mu}+a_{\tau}).
\end{eqnarray}
Using in these formulas the experimental limits from Table II (without brackets) we derive the exclusion curves plotted in Figs. \ref{fig-6}, \ref{fig-7}.
As seen, the limits for low values of  $m_{N}$ in this case are 2 order of magnitude more stringent than in our joint analysis carried out in sec. \ref{sec-4-1} without additional assumptions on the mixing matrix elements. The finite detector size effect is going to  affects these limits only in the region $m_{N}<400$ MeV, weakening them in 2-3 orders of magnitude. This follows from Eq. (\ref{Ueff}).  As we commented in sec. \ref{sec-4-2}, precise impact of this effect depends on the concrete experimental setup. 

\begin{table}[h!]  
\begin{center}
{{\bf Table II:} Experimental and theoretical upper bounds on the branching ratios of meson and $\tau$ decays. The experimental bounds are taken from Ref. \cite{PDG}.
The theoretical upper bounds are shown in the square and curly brackets and are discussed in sec. V. }\\[5mm]
\begin{tabular}{|c|c|c|c|c|c|}\hline
$Br(M \rightarrow \pi l_1l_2  )$& $K$& $D$& $D_s$ & $B$& $B_{c}$\\
\hline
$M^+ \rightarrow \pi^- e^+ e^+  $ & $6.4 \cdot 10^{-10}$ [$1.4 \cdot 10^{-8}$]& $3.6 \cdot 10^{-6}$ [$8.6\cdot 10^{-11}$] 
&$ 6.9 \cdot  10^{-4}$ [$1.0\cdot 10^{-9}$] &$ 1.6 \cdot  10^{-6}$ [$2.3\cdot 10^{-13}$] & \{$4.1\cdot  10^{-11}$\} \\
\hline
$M^+ \rightarrow \pi^- \mu^+ \mu^+  $  & $3  \cdot  10^{-9}$ & $4.8  \cdot  10^{-6} \{3.1\cdot  10^{-7}\}$ &$ 2.9 \cdot  10^{-5} \{3.7\cdot  10^{-6}\}$ &$ 1.4  \cdot  10^{-6}     \{3.4\cdot  10^{-10}\}$ &\{$5.4\cdot  10^{-8}$\} \\
\hline
$M^+ \rightarrow \pi^+ \mu^- e^+  $ &$1.3 \cdot  10^{-11}$ [$7.2\cdot 10^{-8}$]& $3.4  \cdot  10^{-5}$ [$2.5\cdot 10^{-10}$] &$ 6.1 \cdot  10^{-4}$ [$3.1\cdot 10^{-9}$] &$ 1.7 \cdot  10^{-7}$ [$5.7\cdot 10^{-13}$]& \{$1.0\cdot  10^{-10}$\} \\
\hline
$M^+ \rightarrow \pi^- \mu^+ e^+  $  & $5  \cdot  10^{-10}$ [$6.9\cdot 10^{-8}$] & $5 \cdot  10^{-5}$ [$1.7\cdot 10^{-10}$] &$ 7.3 \cdot  10^{-4}$ [$2.2\cdot 10^{-9}$] &$ 1.3  \cdot  10^{-6}$ [$4.6\cdot 10^{-13}$]&\{ $8.3\cdot  10^{-11}$\}\\
\hline
$M^+ \rightarrow \pi^+ \tau^-e^+  $ & - &  - & \{$7.1\cdot  10^{-10}$\} & \{$8.4\cdot  10^{-12}$\} & \{$1.4\cdot  10^{-9}$\} \\
\hline
$M^+ \rightarrow \pi^- \tau^+ e^+  $ & - &  - & \{$7.1\cdot  10^{-10}$\} & \{$8.4\cdot  10^{-12}$\} & \{$1.4\cdot  10^{-9}$\} \\
\hline
$M^+ \rightarrow \pi^+ \tau^- \mu^+  $ & - &  - &- & \{$2.5\cdot  10^{-9}$\} &\{$4.3\cdot  10^{-7}$\} \\
\hline
$M^+ \rightarrow \pi^ -\tau^+ \mu^+  $ & - &  - & -& \{$2.0\cdot  10^{-9}$\} & \{$3.5 \cdot  10^{-7}$\} \\
\hline
$M^+ \rightarrow K^- e^+ e^+  $ & - &  \{$4.0\cdot  10^{-12}$\} & \{$5.3\cdot  10^{-11}$\}  
& \{$1.6\cdot  10^{-14}$\} 
& \{$3.1\cdot  10^{-12}$\}\\
\hline
$M^+ \rightarrow K^+ \mu^- e^+  $ &-& \{$9.8\cdot  10^{-12}$\} &  \{$1.3 \cdot  10^{10}$\} 
& \{$4.2\cdot  10^{-14}$\} 
& \{$7.8\cdot  10^{-12}$\} \\
\hline
$M^+ \rightarrow K^- \mu^+ e^+  $ &-& \{$8.0\cdot  10^{-12}$\} & \{$1.0 \cdot  10^{-10}$\}  & \{$3.4 \cdot  10^{-14}$\} & 
\{$6.3\cdot  10^{-12}$\} \\
\hline
$M^+ \rightarrow K^- \mu^+ \mu^+  $ & - &  \{$7.7\cdot  10^{-9}$\} & \{$1.0\cdot  10^{-7}$\} & \{$1.7\cdot  10^{-11}$\} & 
\{$2.8\cdot  10^{-9}$\}\\
\hline
$M^+ \rightarrow K^+ \tau^{-}e^{+}  $ & - &  - & -& \{$9.2 \cdot  10^{-14}$\} & \{$1.6\cdot  10^{-11}$\}  \\
\hline
$M^+ \rightarrow K^- \tau^{+}e^{+}  $ & - &  - & -&\{$9.2 \cdot  10^{-14}$\}  &\{$1.6\cdot  10^{-11}$\}  \\
\hline
$M^+ \rightarrow K^+ \tau^{-}\mu^{+}  $ & - &  - & -& \{$1.4\cdot  10^{-10}$\} & \{$2.4\cdot  10^{-8}$\} \\
\hline
$M^+ \rightarrow K^- \tau^{+}\mu^{+}  $ & - &  - & -& \{$1.2\cdot  10^{-10}$\} &\{$2.1\cdot  10^{-8}$\} \\
\hline
\hline
$Br(\tau^- \rightarrow l^{\mp} \pi^{\pm} \pi^{-})  $ & $1.2\cdot  10^{-7} \ (e^{-})$ &  $2.0\cdot  10^{-7} \ (e^{+})$ & $2.9\cdot  10^{-7} \ (\mu^{-})$& 
$7 \cdot  10^{-8} \ (\mu^{+})$ & \\
\hline
\end{tabular}
\end{center}
\end{table}

\section{Limits: Compatibility and Interplay}
\label{sec-5}

Let us study some implications of the interplay and compatibility of the constraints on the same combinations of parameters derived from different processes.
\subsection{Impact of $0\nu\beta\beta$-decay limits.}
\label{nudbd}
The exclusion curve from neutrinoless double beta decay ($0\nu\beta\beta$) in Fig. \ref{fig-3},  despite the presence of 
some uncertainty in nuclear matrix elements (of a factor $\sim 2$), is so stringent that within this uncertainty it overrides all the other constraints except for the narrow region where the exclusion curve of the beam dump PS 191 experiment \cite{PS191} goes lower.  However the PS 191 $90\%$C.L.  curve was obtained on the basis of various additional assumptions, touching upon both data processing and their theoretical interpretation,  making the corresponding limits not too firm to compete with the $0\nu\beta\beta$-limits. In particular, the PS 191 limits can be evaded if one admits that the sterile neutrinos have dominant decay channels into invisible particles \cite{Atre:2009rg,evade}.    

Let us examine possible implications of the $0\nu\beta\beta$ constraints in the context of our study.
Presently the best experimental lower
bound on the
$0\nu\beta\beta$-decay half life is obtained for ${}^{76}$Ge \cite{KlapdorKleingrothaus:2000sn}:
\begin{equation}\label{H-M}
T^{0\nu}_{1/2}(^{76}Ge) \geq 1.9\times 10^{25}\mbox{yrs}.
\end{equation}
In Ref. ~\cite{Benes:2005hn} this bound was used to constrain
the contribution of Majorana neutrinos of arbitrary mass. In a good approximation this constrain reads:
\begin{eqnarray}\label{dbd-1}
\sum_{k}
\frac{\left |U_{e k}\right|^{2} e^{i\alpha_k} m_{\nu k}}{m_{\nu k}^{2} + q_0^{2}}\leq 5\times 10^{-8} \,\textrm{GeV}^{-1}.
\end{eqnarray}
with $q_0 = 105$ MeV ~\cite{Benes:2005hn}. Here $\alpha_k$ denotes the Majorana phase of the Majorana neutrino state $\nu_k$ of mass $m_{\nu k}$. For the light-heavy neutrino scenario one has:
\begin{eqnarray}\label{dbd-2}
\sum_{N=\textrm{heavy}}
\frac{\left |U_{e N}\right|^{2} }{m_N} e^{i\alpha_N}
+ q_0^{-2}\sum_{i=\textrm{light}}
\left |U_{e i}\right|^{2} e^{i\alpha_i} m_{\nu i}
\leq 5\times 10^{-8} \,\textrm{GeV}^{-1}.
\end{eqnarray}
where $m_N\gg q_0$ and $m_{\nu i}\ll q_0$.
This upper limit, applied to each term separately, leads to a very stringent constraint on the heavy neutrino mass and mixing
\begin{eqnarray}\label{dbd-21}
\frac{\left |U_{e N}\right|^{2} }{m_N}  \leq 5\times 10^{-8} \,\textrm{GeV}^{-1}.
\end{eqnarray}
One may use this limit to constraint LNV and LFV processes with $e^{\pm}$ in the final state.
From Eq. \rf{GT-comp-incl}, with the fact that $a_l(m_N) |U_{l N}|^2 \leq  \Gamma_{N}$, we obtain a useful inequality 
\ba{dobleUb}
{\cal F}_{l_{1}l_{2}}(m_{N}) \equiv \frac{|U_{l_1 N}|^2|U_{l_2 N}|^2}{\Gamma_{N}} \leq  \frac{|U_{l_1 N}|^2}{a_{l_2}} \ \ \ \ l_1,l_2= e,\mu,\tau
\ea
 valid for all values of  $m_N$. Combining Eqs.  \rf{dobleUb} and (\ref{dbd-21}) we get
\ba{dobleUb-1}
{\cal F}_{e l}  \leq   \frac{(m_{N}/1 \mbox{ MeV})}{a_{l}} \cdot 5\times 10^{-11} \ \  \mbox{for} \ \ 
l=e,\mu,\tau.
\ea
Using these limits in Eqs. (\ref{LNV-res})-(\ref{tau-dec}) we can derive upper limits on some LNV and LFV decays imposed by non-observations of $0\nu\beta\beta$-decay (\ref{H-M}).  Varying $m_{N}$ within the resonant regions \rf{domain} of the considered decays 
we determine their largest rates compatible with Eq.  \rf{dobleUb-1}, which we present in Table II in square brackets.  These limits represent absolute upper bounds on given decays rates within the sterile neutrino extension of the SM, compatible with $0\nu\beta\beta$-constraints. Here an important assumption is implied.  
We assumed that all the Majorana phases in Eqs. (\ref{dbd-1}), (\ref{dbd-2}) are trivial $\alpha_{k}=2\pi n$. 
This case corresponds to no cancellation between different terms in Eqs. (\ref{dbd-1}),  (\ref{dbd-2}) when the limit (\ref{dbd-21}) is valid. 
If there are more than one heavy neutrino state $N_k$,  with
different non-trivial Majorana phases $\alpha_N$, then different terms in the first sum of (\ref{dbd-2}) may
compensate each other, reducing the individual
$0\nu\beta\beta$-constrain on each of them.
Note that even with the presence of only one heavy neutrino there
may happen this kind of  compensation between the heavy $N$ and
the three light neutrinos $\nu_i$, since they all coherently contribute to
$0\nu\beta\beta$-decay.  Thus we conclude that observation of some of the processes in Table II, at rates above the indicated limits from $0\nu\beta\beta$, shown in square brackets,  may point to the following:
{\bf (a)} Presence of a sterile Majorana neutrino N with the mass $m_{N}$ in the resonant range of these processes and that the Majorana phases are non-trivial;  {\bf (b)} Presence of a Dirac sterile neutrino in this mass range, if LFV decay is observed; 
{\bf (c)} There is an exotic low mass scale new physics with light particles within the same resonant range of masses, but other than the sterile neutrino. 
So far most of the existing experimental limits shown in Table II are significantly weaker than the corresponding $0\nu\beta\beta$ limits, except for $K\rightarrow ee \pi, \mu e \pi$. We note, however, that the latter is not yet definite and requires more detailed analysis \cite{Atre:2009rg}
of the corresponding experimental data. Uncertainties are related to the fact that  for small mixing only part of the produced heavy 
sterile neutrinos decay in the detector and as a consequence the actual experimental bounds on $K\rightarrow ee \pi, \mu e \pi$ could be
several orders of magnitude lower than indicated in Table II.  
Therefore more careful analysis of the existing experimental data  and further searching for decays with $e^{\pm}$ in final states may provide an important information on the Majorana phases of neutrino mixing matrix. These phases, in general, may lead to CP violation in the neutrino sector.  However the above discussed analysis is unable to distinguish the CP violating values of the Majorana phases from their values $\alpha_{k}= \pi n$ which lead solely to opposite CP parities of some neutrino states which do not lead to CP violation. In both cases there is a cancellation between different terms in Eqs. (\ref{dbd-1}), (\ref{dbd-2}). 
As to the experimental prospects,   it is expected in the future an improvement of the 
experimental bounds on the branching ratios of the considered decays by an order of magnitude or more.
This is, in particular, possible in LHCb experiment. 

\subsection{Limits on as yet experimentally unconstrained decays.}
\label{unconstrained}
Presently there are no experimental limits on semileptonic decays of mesons with  $\tau$ in final states 
\mbox{$M_{1} \rightarrow \tau \ l_2 \ M_{2}$},  and on any semileptonic decay of $B_c$ like $B_c \rightarrow \ l_1 \ l_2 \ M$. Below we derive theoretical upper limits for these decays within the considered sterile neutrino extension of the SM, on the basis of the fact that the maximal rates are reached in the resonant range for the sterile neutrino masses. Although such limits are linked to a specific scenario, they may have a more general meaning,  considering the fact that the sterile neutrino extension of the SM is the only reasonable model having a light particle, sterile neutrino $N$, which can cause resonant enhancement of the processes in question. Other known beyond the SM scenarios are associated with larger mass scales and heavier particles outside the resonant regions of these processes and then their contribution is expected to be much smaller than that from the resonant mass sterile neutrino. 

For derivation of the limits on as yet experimentally unconstrained decays we use the constraint  \rf{dobleUb-1} and 
    \ba{cotaUt}
 {\cal F}_{\tau l} &\leq& \frac{\Gamma^{exp} {(\tau \rightarrow l \ \pi \ \pi)} }{ \pi G^l (z_\tau) m_N}; \ \  \ \ \ \ \ \ \ \ \ \ \ \ \ \ \ \ \ \ \ \ \
 {\cal F}_{l_{i}l_{j}}  \leq \frac{\Gamma^{exp} {(M \rightarrow l_i l_j \ \pi)} }{ \pi \kappa_{ij} (G_M^{ij} (z_M)+G_M^{ji} (z_M)) m_N};\\
 \nn
 {\cal F}_{\mu l} &\leq&  \frac{|U_{\mu N}|^{2}}{a_{l}} \ \ \ \  \mbox{for} \ \  l=\mu,\tau;\ \ \ \ \ \ \ \ \ \ \ \ \
 |U_{\tau N}|^{2} \leq 0.018  \ \ \ \  \mbox{for} \ \ \ \  m_{N} \geq m_{\tau}.
 \ea  
Here we used notations of Eq. \rf{dobleUb}. The first two constraints are derived from Eqs.  (\ref{LNV-res})-(\ref{tau-dec}) and are 
valid in the range $ m_{\pi} + m_{l} \leq m_N \leq m_\tau - m_\pi$ for $\tau$-decays and for meson decays in 
the ranges indicated in Eqs.  \rf{domain}. The third constraint is a particular case of the inequality \rf{dobleUb}. The last one originates from precision electro-weak measurements \cite{Almeida:2000pz, Bergmann}.
In order to derive upper limits on as yet unconstrained decays we scan their corresponding resonant ranges \rf{domain},  selecting for each value of $m_{N}$ the largest value of their rates compatible with the constraints in Eqs. \rf{dobleUb-1} and \rf{cotaUt}. 
In Eqs. \rf{cotaUt} we use the experimental upper bounds  
$\Gamma^{exp} (\tau \rightarrow l \ \pi \ \pi) $, $\Gamma^{exp} (M \rightarrow l_i l_j \ \pi)$ shown in Table II. In the third constraint in Eq.  \rf{cotaUt}
for $|U_{\mu N}|^2$ we select the most stringent of  the known (see Fig. \ref{fig-4}) limits for a given value of $m_{N}$.
The resulting upper limits on the branching ratios for some LFV and LNV processes are listed in Table II in curly brackets.  Here,  for completeness, we also presented limits derived in this way for several  decays already constrained from experiment.
Note that the limits presented for the decays with $e^{\pm}$ in the final states imply trivial Majorana phases in the neutrino sector discussed in section \ref{nudbd}.
As seen from Table II  only the experiments searching for the LNV and LFV decays of $D$ and $D_{s}$ are rather close to probe their rates not excluded by our analysis.  Searches for other decays are far away from this perspective. 
As to $K$-meson LFV and LNV decays, we refer to our comments at the end of section \ref{nudbd}.
\section{Summary and Conclusions}
\label{sec-6}
We have studied LFV and LNV decays of $\tau$ and $B, D, K$ mesons  mediated by heavy sterile
neutrinos. We focussed on the dominant mechanism via resonant neutrino contribution.

On the basis of a joint analysis of experimental bounds  on LNV and LFV decays listed in Table II, we extracted upper limits on the heavy sterile  neutrino mass $m_{N}$ and mixing $|U_{eN}|, |U_{\mu N}|, 
|U_{e N} U_{\mu N}|,  |U_{e N} U_{\tau N}|,
|U_{\mu N} U_{\tau N}|$ in a model independent way, without additional assumptions on these parameters. Our limits derived from the above mentioned decays are shown in Figs. \ref{fig-3}-\ref{fig-5}, and represent the best possible limits which can be extracted in this way from the existing experimental data on the studied decays. 

We also derived the limits on the sterile neutrino mass and mixing with an ad hoc assumption 
$|U_{e N}|\sim |U_{\mu N}|\sim |U_{\tau N}|$ frequently used in the literature. The resulting limits, as expected, turned out to be much more stringent than those we obtained  in a model independent way. This illustrates a significant impact of such assumptions on the extracted limits  making such limits  model dependent.

In our analysis we used the method of calculation of the total decay rate of heavy sterile neutrino based on an
inclusive approach. Instead of a detailed calculation of the sum of  all the possible decay channels with different hadrons in final state, we approximated hadronic final states, starting from certain mass threshold, with $q\bar{q}$-pair,  as  suggested by Bloom-Gilman duality. This allowed us to avoid theoretical uncertainties related with the decay constants of heavy mesons and take into account in average all the decay channels, some of which cannot be included in the channel-by-channel approach.  

Special emphasis has been made on the stringent limits on the heavy Majorana neutrino mass and mixing from non-observation  of 
$0\nu\beta\beta$-decay. In particular, we used these limits to derive indirect limits on various LNV and LFV processes. 
Our results, displayed in Table II,  have been obtained assuming 
zero Majorana phases in the neutrino mixing matrix.
As seen from Table II, our expectations for  observability of the decays with both $e^{\pm}e^{\pm}$ 
and $e^{\pm}e^{\mp}$ in the final states are pessimistic.  
One of the messages of the present paper is that despite of such discouraging prediction these processes are worth searching.
The point is that any observation of such decays above the limits set by the $0\nu\beta\beta$-decay constraints in Eq. (\ref{dbd-21}) would most likely  point to the existence of heavy Majorana sterile neutrino in the resonant regions of these processes, and non-trivial Majorana phases, as explained in sec. \ref{nudbd}.  

We applied our and other existing limits on heavy sterile neutrino mass $m_{N}$ and mixing $U_{lN}$ for the prediction of upper bounds on the  rates of LFV and LNV decays of $B$ and $D$ mesons, some of which are as yet experimentally unconstrained. 
These limits are shown in Table II, and indicate that only experiments searching for LNV and LFV decays of $D$ and $D_{s}$ are rather close to probe the rates of these processes, not excluded by our analysis.  Searches for other decays are far away from this perspective. \\[5mm]
{\bf Acknowledgements} This work is supported by FONDECYT (Chile) under projects 1100582, 1100287, 
\mbox{CONICYT(Chile)} under project 791100017   
and Centro-Cient\'\i fico-Tecnol\'{o}gico de Valpara\'\i so PBCT ACT-028.

\end{document}